\documentclass[preprint,12pt]{elsarticle}

\usepackage{amssymb}
\usepackage{amsmath}
\usepackage{lineno}
\usepackage{hyperref}
\usepackage{diagbox}
\usepackage{siunitx} 

\journal{NIM A}

\begin{document}

\begin{frontmatter}

\title{Performance of the Endcap Time-of-Flight detector in the STAR beam-energy scan}

\author[ucd]{Mathias C. Labonté\corref{cor1}}
\author[ucd]{Daniel Cebra}
\author[ucd]{Zachary Sweger}
\author[rice]{Geary Eppley}
\author[rice]{Frank Geurts}
\author[heidelberg]{Yannick Söhngen }
\author[heidelberg]{Norbert Herrmann}
\author[heidelberg]{Esteban Rubio}
\author[heidelberg]{Philipp Weidenkaff}
\author[heidelberg,gsi]{Ingo Deppner}
\author[gsi]{Pierre-Alain Loizeau}
\author[gsi]{Jochen Frühauf}
\author[gsi]{David Emschermann}
\author[gsi]{Florian Seck}
\author[cre]{David Tlusty}
\author[ustc]{Dongdong Hu}
\author[ustc]{Yongjie Sun}
\author[rice]{Tonko Ljubicic} 
\author[tshu]{Yi Wang}

\address[ucd]{Department of Physics and Astronomy, University of California, Davis, Davis, CA 95616, USA}

\address[rice]{Department of Physics and Astronomy, Rice University, 6100 Main St., Houston, TX 77005-1827, USA}

\address[heidelberg]{Physikalisches Institut, Im Neuenheimer Feld 226, 69120 Heidelberg, Germany}

\address[gsi]{GSI Helmholtzzentrum für Schwerionenforschung GmbH, Planckstr. 1, 64291 Darmstadt, Germany}

\address[ustc]{University of Science and Technology of China, Hefei, Anhui 230026, China}

\address[tshu]{Tsinghua University, Beijing 100084, China}

\address[cre]{Creighton University, Omaha, Nebraska 68178, USA}

\cortext[cor1]{Corresponding author. Email: mlabonte@ucdavis.edu}

\begin{abstract}

The STAR experiment at RHIC at Brookhaven National Laboratory completed the installation of an endcap time-of-flight subsystem in February 2019.  The subsystem provided essential mid-rapidity particle identification  for the fixed-target portion of phase II of the STAR beam energy scan.  The fixed-target program allowed the experiment to access center-of-mass energies from $\sqrt{s_{_{NN}}} = 3.0$ GeV to $\sqrt{s_{_{NN}}} = 7.7$ GeV, not accessible by colliding beams.  The system's detectors and readout electronics were designed for the CBM experiment at FAIR and adapted for use at STAR.  In this paper, we describe the details of the system in terms of geometrical layout, acceptance, calibration, hit reconstruction, and particle identification. The system achieved a time resolution of about 70 ps and a track-matching efficiency of about 70\%, meeting the design goals of the project.   
\end{abstract}

%
%
%


\end{frontmatter}  

\section{Introduction}
\label{intro}
Studying the phase diagram of nuclear matter under extreme conditions is the main focus of the beam energy scan ~(BES) at the Relativistic Heavy-Ion Collider ~(RHIC) at Brookhaven National Laboratory \cite{BESWHITE}.  RHIC collided heavy nuclei ($^{197}\rm Au$) at a range of center-of-mass energies to study the properties of nuclear matter at different temperatures and baryon chemical potentials ($\mu_B$). Achieving low-energy collisions required injecting beams into the RHIC synchrotrons at energies below the design injection energies (9.8 GeV for each ring, $\sqrt{s_{_{NN}}} = 19.6$ GeV). Circulating beams below the design energy of the synchrotons challenged the ability of RHIC to focus the ion beams. The lowest collision energy achieved in the collider configuration was $\sqrt{s_{_{NN}}} = 7.7 \rm \ GeV$ ~(corresponding to $\mu_B \approx$ \SI{420}{\mega\electronvolt} \cite{STAR:2017sal}). In order to more extensively scan the quantum chromo-dynamic phase diagram, it was possible to further lower the center of mass energies at RHIC by implementing a fixed-target (FXT) configuration. The FXT configuration was implemented by STAR ahead of BES phase II (BES-II) by installing a gold foil in the beam pipe near the west end of the detector and circulating beam in only one of the two RHIC synchroton rings. The FXT configuration allowed for collisions as low as $\sqrt{s_{_{NN}}} = ~3.0 \rm \ ~GeV$ ~($\mu_B \approx $ 720 MeV). STAR recorded data at 12 collision energies in the FXT configuration ($\sqrt{s_{_{NN}}}  = 3.0, \ 3.2, \ 3.5, \ 3.9, \ 4.5, \ 5.2, \ 6.2,  \ 7.2, \ 7.7, \ 9.2, \ 11.5, \ 13.7$ GeV), with the four highest energies overlapping with data taken in the collider mode allowing for validation of the new experimental geometry.

\begin{figure*}[h!t!b]
    \centering
    \includegraphics[width = 0.9\textwidth]{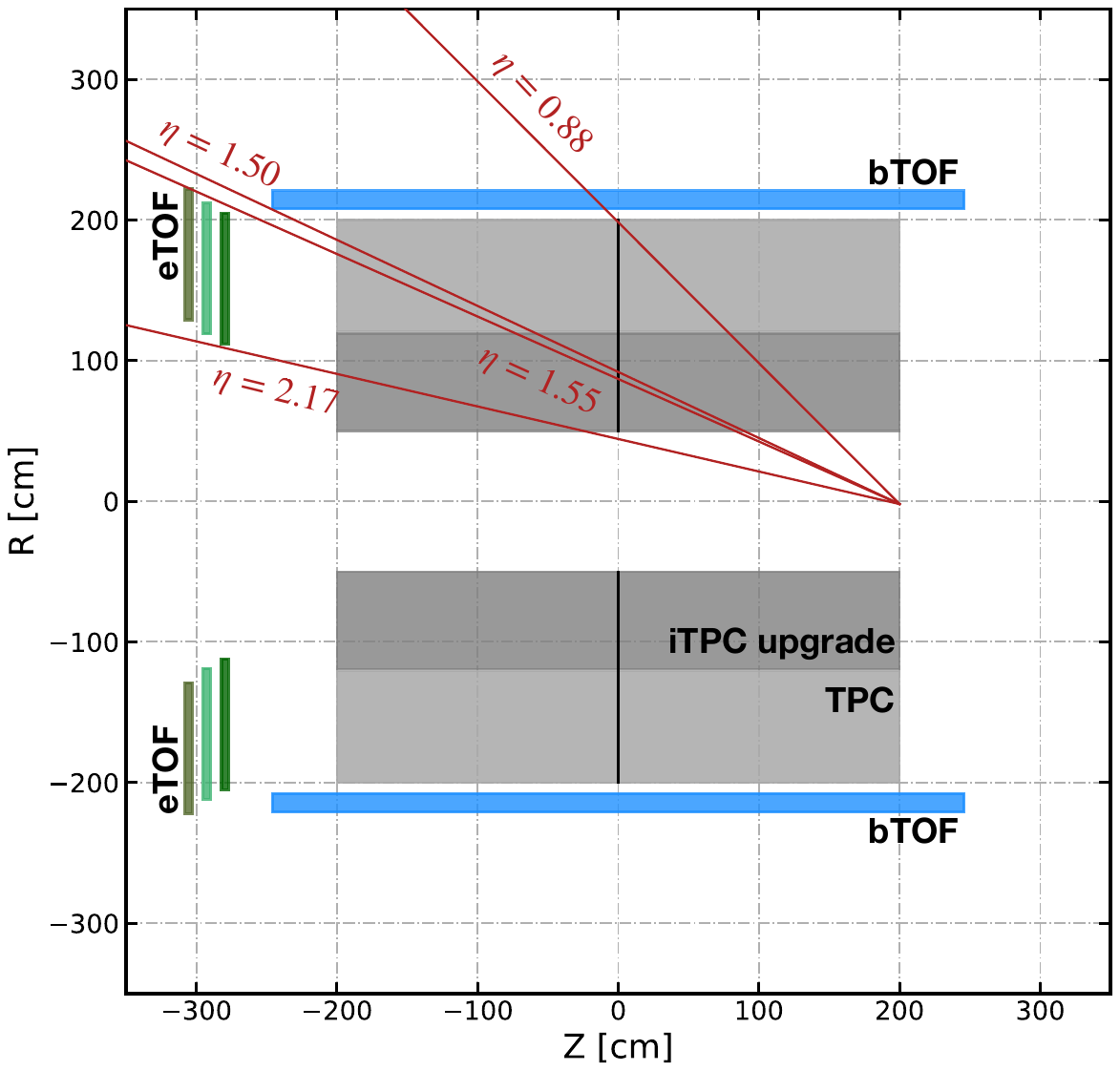}
    \caption{Schematic of the STAR central barrel. The TPC provides the momentum measurement necessary for TOF PID. The TPC is a cylinder with radius of 200 cm and a length of 200 cm for each half. The 
    iTPC \cite{BESupgrades, iTPCRelated} is also an upgrade for BES-II, which increased the low-$p_T$ coverage and tracking resolution \cite{star2015technical}. eTOF is shown in green with different shades corresponding to the three layers of the eTOF wheel. The inner and outermost radii of eTOF are 112 cm and 222 cm, respectively. The central high voltage cathode of the STAR TPC is represented by the line at $Z = 0$ cm. Red lines indicate pseudorapidity ($\eta$) values in FXT mode at key points in the STAR geometry.}
    \label{fig:etof_yz}
\end{figure*}

The FXT program comes with its own challenges for a collider detector like STAR \cite{Star:2002eio}.   Midrapidity moves outside of the acceptance of the barrel time-of-flight detector (bTOF) \cite{Llope:2012zz} for collision energies higher than $\sqrt{s_{_{NN}}}  = 3.9$ GeV ($0 < \eta_{_{\rm{bTOF}}} < 1.5$ in FXT mode). To expand the available phase-space and recover midrapidity for the higher FXT energies, an endcap time-of-flight detector (eTOF) 
\cite{STAR:2016gpu} was installed in 2019 for BES-II. The inclusion of eTOF extended the pseudorapidity coverage by 0.7 units ($1.55 < \eta_{_{\rm{eTOF}}} < 2.17$)
in FXT mode, and allowed for midrapidity measurements for collision energies up to $\sqrt{s_{_{NN}}} = 7.7$ GeV.  Figure~\ref{fig:etof_yz} shows a cartoon schematic of the STAR central barrel, with eTOF located between $z  =$ -303 to -270 cm. Pseudorapidity in FXT mode is defined in reference to the target, which is located at $z = 200$ cm. The other main detectors in the STAR central barrel are also shown, the time projection chamber (TPC) \cite{TPC1} and
bTOF. The TPC provides tracking capability as well as low-momentum particle identification through measurements of energy loss ($dE/dx$). Time of flight detectors measure a particle's mass:
    \begin{equation}
    \label{TOF}
        m = \frac{p}{c} \cdot \sqrt{ \frac{c^2 \cdot \Delta t^2}{\ell^2} -1 } 
    \end{equation}
    where $p$ is the particle's momentum, $c$ the speed of light, $\Delta t$ the time of flight and $\ell$ the path length from the primary vertex to the intersection on the counter surface. Here, $p$ and $\ell$ are provided by the TPC particle tracking.
In this paper, the performance and capabilities of eTOF during BES-II will be discussed as well as its utility for physics analysis at STAR.

\section{The endcap time-of-flight detector}

\begin{figure*}[ht]

    \centering
    \includegraphics[width=0.43\textwidth]{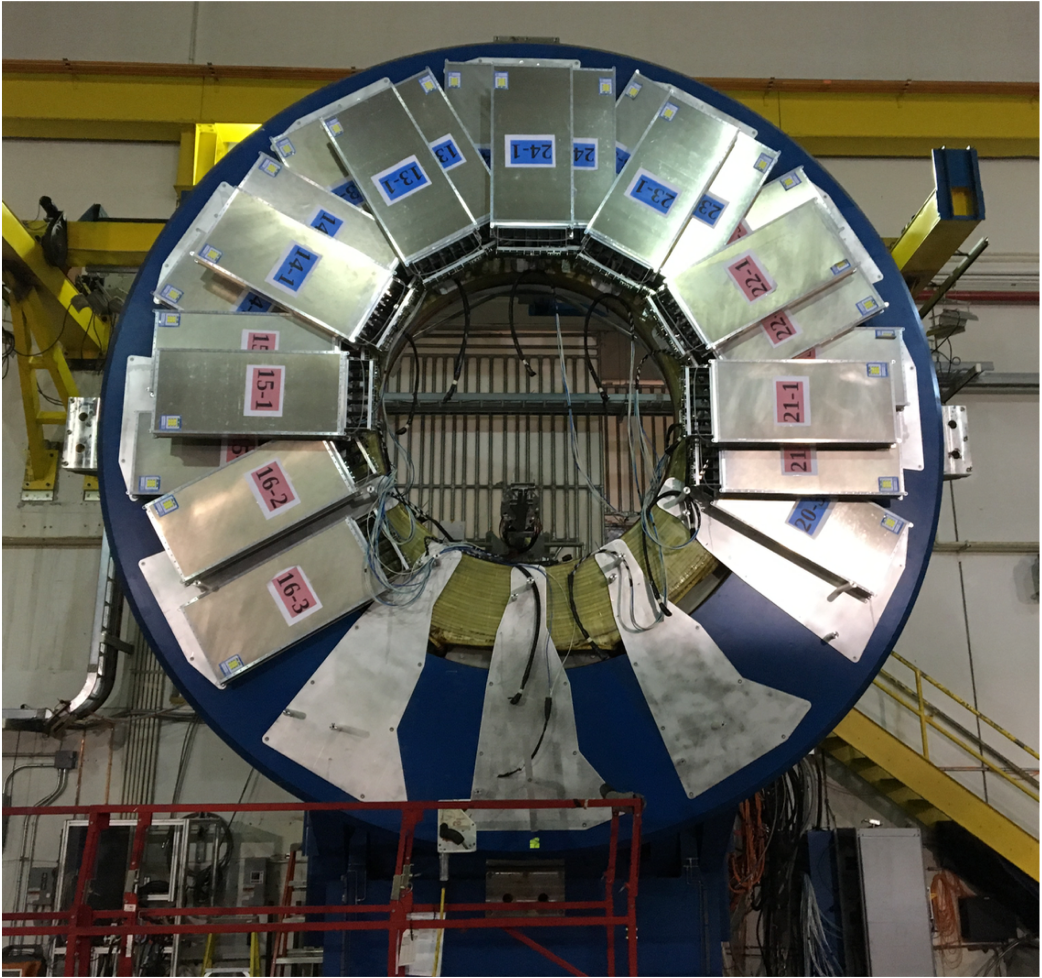}
    \includegraphics[width=0.5\textwidth]{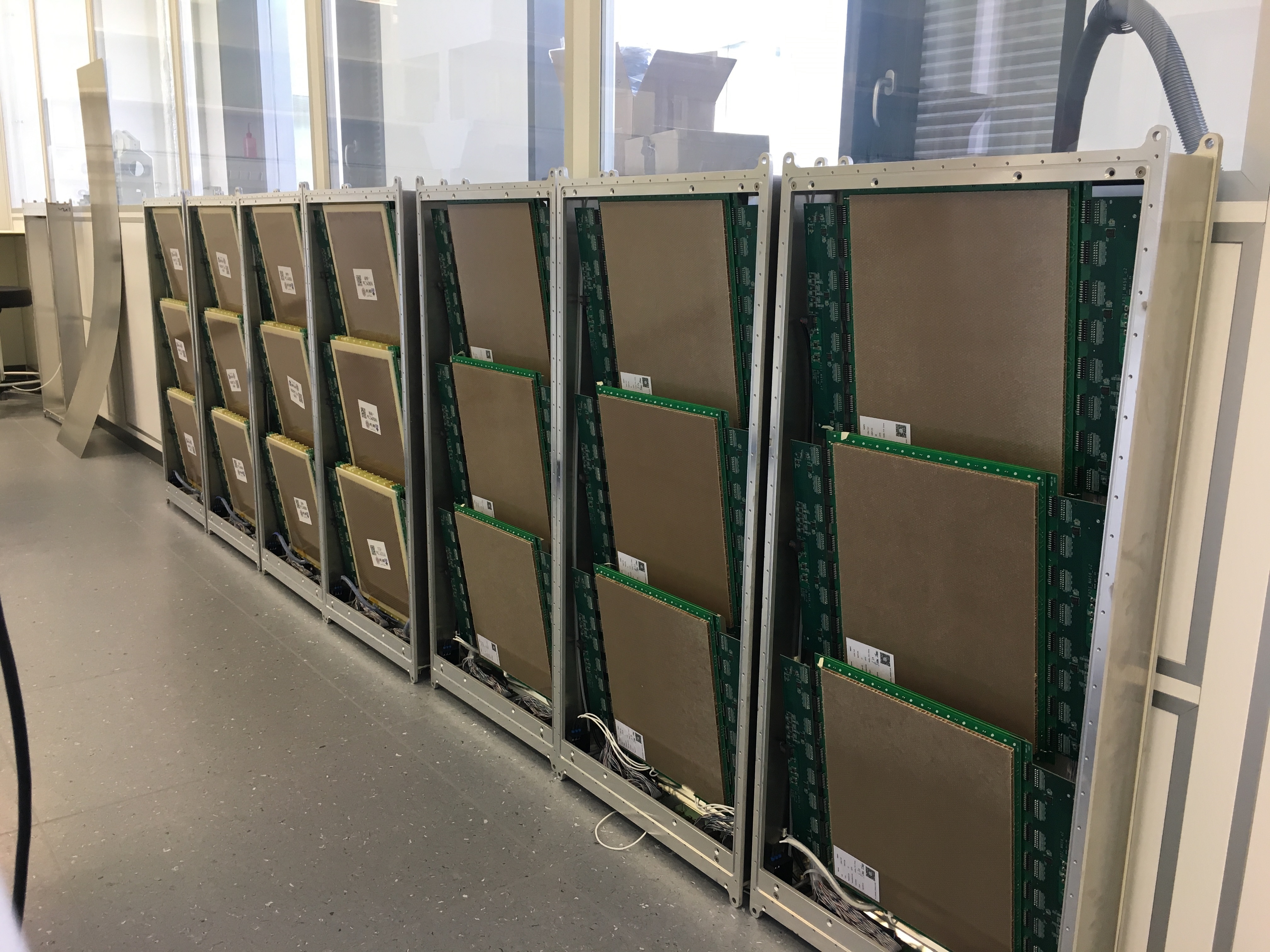}
    \caption{Left: Photograph of the eTOF wheel (not fully installed). Modules are arranged in three layers. Right: open modules in the clean room equipped with 3 MRPCs each.}
    \label{fig:etof}
    \end{figure*}

\begin{figure*}[ht]

    \centering
    \includegraphics[width=\textwidth]{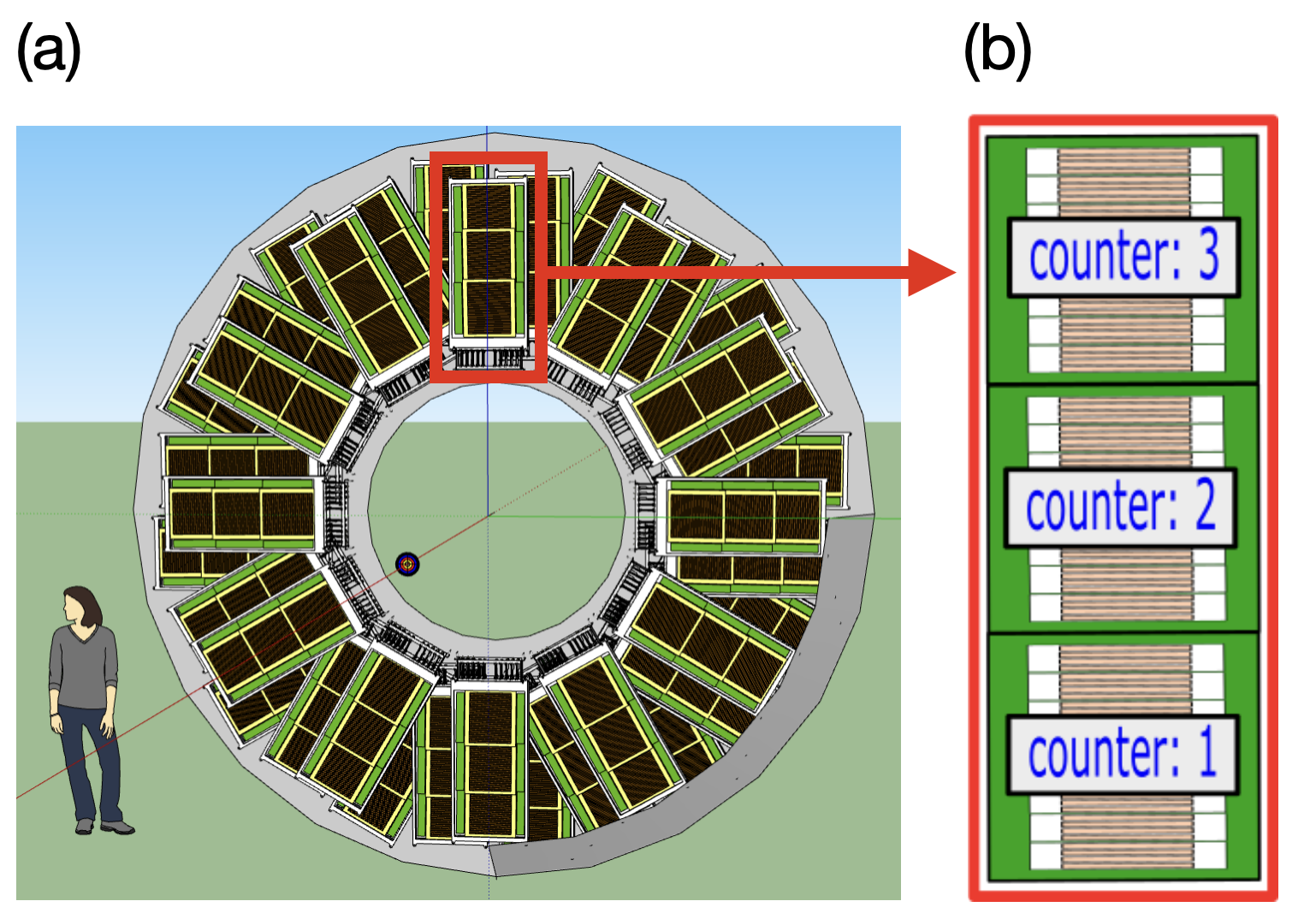}
    \caption{Schematic layout of the eTOF detector. (a): full wheel structure composed of 12 sectors. (b): single module containing 3 MRPC's. }
    \label{fig:etof_schem}
    \end{figure*}

    The endcap-time-of-flight detector is composed of 108 multi-gap resistive plate chambers (MRPCs \cite{Fonte:1999iw}) arranged in 36 modules with a total of 6912 read out channels. Each module has an active area of 92~cm $\times$ 27~cm. There are two, 2~cm overlap regions of the 3 MRPCs in a module. The total size of the gas-tight module box is 120~cm $\times$ 49~cm $\times$ 11~cm, including the crate for the readout electronics at one end. Three modules form an eTOF sector. There are 12 sectors numbered 13 to 24 with a rotation angle of 30$^\circ$ per sector that form a wheel-like structure matching sectors 13 to 24 of the east side of the TPC. Due to the overlap of the modules within one sector, 3 layers are obtained with distances of 277, 290, and 303 cm, respectively, from the center of the TPC. The radial extension of the active area ranges from 112 to 222 cm from the center of the beam. The eTOF modules were operated with a gas mixture of 80 parts Freon R134A, 4 parts Isobutane, 1 part SF6. Installed eTOF modules are shown on the left side of Fig.~\ref{fig:etof} and Fig.~\ref{fig:etof_schem}.

    \begin{figure*}[ht]
    \centering
    \includegraphics[width=0.7\textwidth]{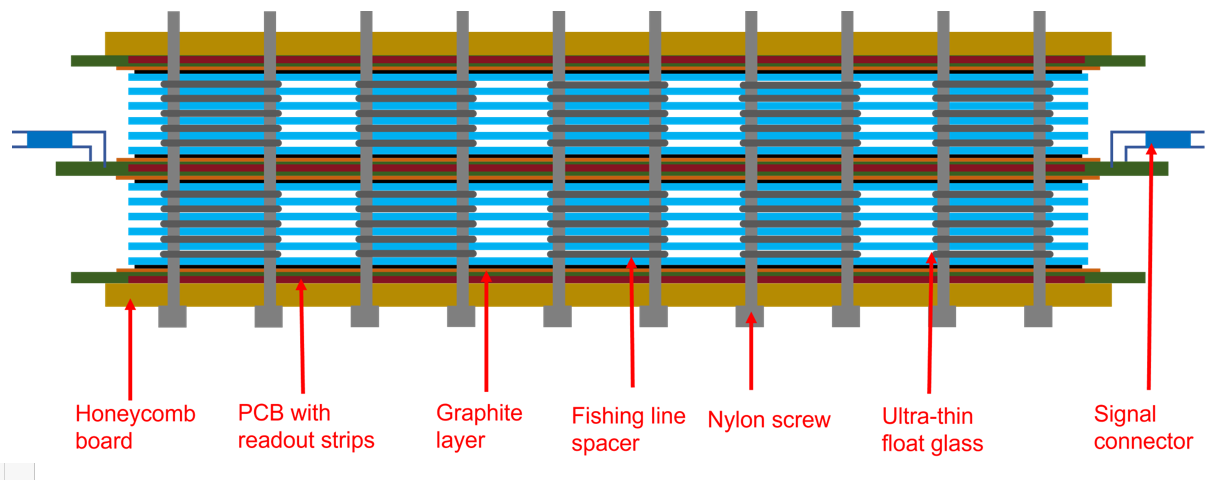}
    \caption{Scematic drawing of the MRPC3b used at eTOF. }
    \label{fig:mrpc3b}
    \end{figure*}

    The 3 MRPCs within one module are tilted toward the beam axis by 10$^\circ$ (cf. right side of Fig.~\ref{fig:etof}) and overlap with their neighboring detectors by almost 2 readout strips. The MRPCs themselves have a double-stack configuration and encase a gas volume subdivided by multiple resistive plates between the high-voltage electrodes and read-out strips (see Fig.~\ref{fig:mrpc3b}). 
    Two types of MRPCs originally designed for two regions with different rate exposure within the CBM time-of-flight wall \cite{Deppner:2012zz, Deppner:2018kcf} are used. An MRPC (called MRPC3a) developed by Tsinghua University in Beijing (THU) can handle charged particle fluxes of up to several tens of kHz/cm$^2$. They use low-resistive glass plates, $\approx 10^{10} \Omega /$cm. For more details, see \cite{Wang:2001kzg}. Modules equipped with this MRPC type cover eTOF sectors 15, 16, 21 and 22. The other eight sectors are equipped with counters (called MRPC3b) containing float-glass plates, $\approx 10^{12} \Omega / cm$, built at the University of Science and Technology China in Hefei (USTC). The MRPC3b can handle rates up to 2.5 kHz/cm$^2$ (cf. \cite{Wang:2023vqi, Zhou:2019kew, Hu:2018rqg}). More technical details are summarized in Tab. \ref{tab:tech_detais}.

\begin{table*}[h!]
\centering
\noindent
\begin{tabular}{|l||*{2}{c|}}\hline
\diagbox{Technical parameter}{MRPC type}
&\makebox[3em]{MRPC3a}&\makebox[3em]{MRP3b}\\\hline\hline
Produced at &Tsinghua Uni. (THU)& USTC\\\hline
Number of gaps &2 $\times$ 4&2 $\times$ 5\\\hline
Gap size &0.25 mm& 0.23 mm\\\hline
Glass type& low resistive& float glass\\\hline
Glass thickness &0.7 mm& 0.28 mm\\\hline
Number of readout strips &32&32\\\hline
Pitch &1 cm&1 cm\\\hline
Active strip length $\times$ strip width & 26.8 cm$\times$ 7 mm& 27 cm $\times$ 7 mm\\\hline
\end{tabular}
\caption{Technical information for the MRPC detectors.}
\label{tab:tech_detais}
\end{table*}

   Both sides of the readout strips are connected to the front-end cards which sit close to the detector inside the module box. Each PADI FEE card contains four, 8-channel PADI-X preamplifier and discriminator ASICs \cite{Ciobanu:2021qlz}. The PADI sends LVDS signals via twisted-pair cables and a feed-through PCB to the GET4 time-to-digital (TDC) ASIC \cite{Deppe:2009rpa}. The GET4 records the arrival time and the time over threshold. The time over threshold is used to correct the arrival time for rise-time effects, a process known as slewing or time-walk correction.
   Each GET4 FEE card contains eight, 4-channel GET4 ASICs and sits in a crate mounted directly to the end of the module box. Six GET4 FEE cards are plugged into one module.  
   The digitized information of the 24 input links (3 GET4 FEE cards with 8 GET4 ASICs each) is collected and serialized in the read-out-board (ROB).  The ROB uses the CERN GBTx ASIC and sends the data  via optical fiber to the FPGA based AFCK board. For one module, 2 ROBs are needed, one for each end of the 96 read-out strips. 
   To keep the system synchronized, the ROBs are supplied with a phase-stable clock of 40 MHz locked to the STAR bTOF clock. The GBTx provides a phase-locked 160 MHz clock to the GET4s which is the reference for the signal time measurements. The clock cycle has a period of 6.25 ns and the GET4 TDC has 112 bins with bin width 55.8 ps. There is no significant non-linearity.
   
   A synchronization signal (sync) with a frequency of 0.86 Hz is used by the GET4s \cite{deppe_flemming_cbm-collaboration_2010} to compare their internal clock counters to the expected value.
   The AFCK board has 6 optical input links and one optical output link and acts as a data concentrator board for one entire sector (2 ROBs x 3 modules). Twelve AFCK boards, installed in a $\mu$-TCA crate and located in the experimental area, send their data via  optical output links to 2 FPGA-based PCIe-cards installed in the eTOF data acquisition server (DAQ-PC) located in the STAR DAQ-room. The eTOF DAQ-PC transfers the processed data to the STAR DAQ system.

    \begin{figure*}[ht] 
    \centering
    \includegraphics[width=0.98\textwidth]{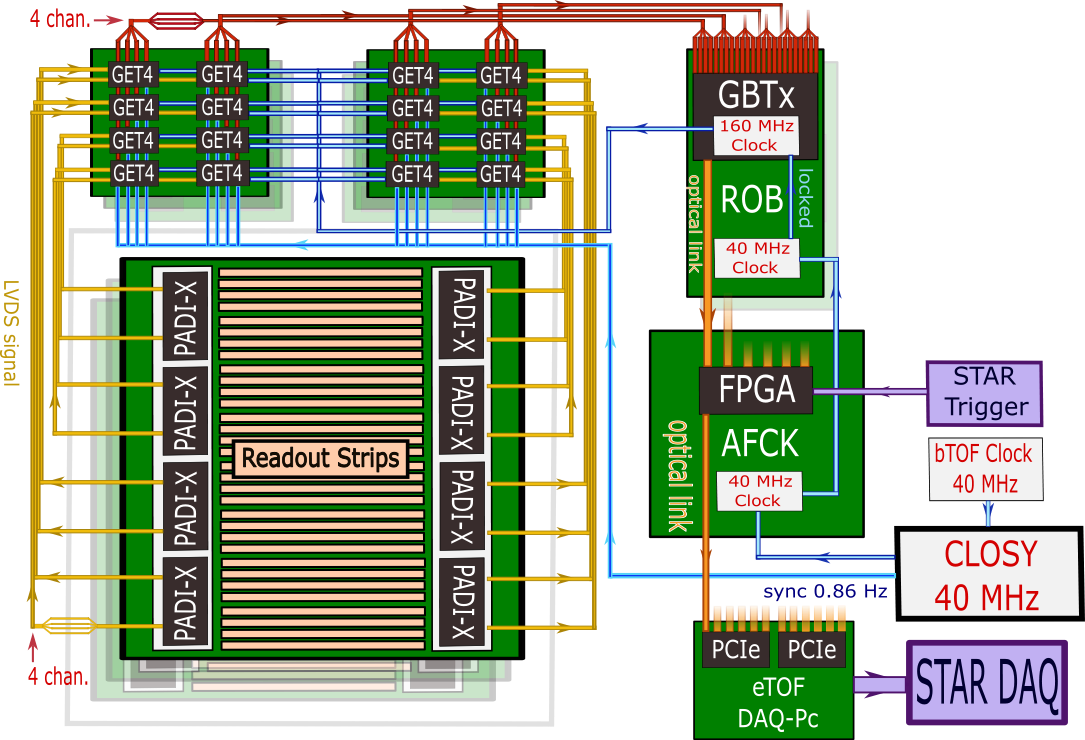}
    \caption{Schematic depiction of the eTOF readout chain and its components. }
    \label{fig:readout}
    \end{figure*}

   The GET4 ASIC is triggerless, by design. As a result, all hits recorded by the eTOF detectors are continuously sent to eTOF DAQ. In the eTOF DAQ PC, eTOF subevent data are produced using an early version of the CBM data acquisition chain based on Flesnet \cite{Cuvelant2023CBM} and the FairMQ framework \cite{fairmq}. The time-stamped messages generated in the firmware running in the AFCK boards are combined by the Flesnet software in units of time slices that hold the data of all channels which have sent a signal in a given time interval. These time slices are then delivered as input to parallel running FairMQ event builder devices 
   which scan the incoming data stream for STAR trigger tokens.
   Tokens are special messages that are injected into the data stream by the AFCK boards and code a signal received from the STAR trigger system (see Fig.~\ref{fig:readout}).    
   Upon finding a STAR trigger token in the data stream, an event is generated and filled with all time stamps within a window of approximately 3 $\mu$s with respect to the trigger token time.
    In the same process, time-stamped status messages describing the operational state (ON / OFF) of each GET4 digitizer are inspected, and a status flag mask is updated to describe the actual state of the system at any moment in time. The actual state of the system is stored in a ``Get4Active'' mask in the eTOF event header.  The subevents are then sent to a single trigger handler process that transmits the data via 1 Gbs ethernet to the STAR DAQ system.

\subsection{Hit reconstruction and track matching}

    STAR data are collected in units of runs generated under stable trigger and DAQ conditions with a time interval of about an hour or less.
    The eTOF data analysis starts with the conversion of raw data acquisition messages to ``digi'' objects that represent 
    the smallest addressable eTOF data item in the STAR/CBM data analysis environment. Digis contain the signal's leading-edge arrival time at the GET4 digitizer, the time over threshold (ToT), and address information in terms of module number, MRPC number, readout strip number, and side of the strip.     

    Two eTOF digis from opposite sides of a strip are combined to form a hit that can be assigned to a TPC track. 
    With momentum and path length information provided by the track and the time of flight originating from the hit, the mass of the detected particle can be obtained according to Eq.~\ref{TOF}, with the time of flight given by the difference of a particle's arrival or stop time and the event's start time provided by bTOF. The bTOF start time is calculated using well-identified pions and protons from the TPC dE/dx measurement and the stop time measured by bTOF. The start time is calculated for each track. Averaging gives the event start time with a resolution of about 21 ps in $\sqrt{s_{NN}}=4.5$ GeV FXT.

    For each digi pair on a given strip, the hit time is the average of the two digi times. The local-$y$ position

    \begin{equation}
        \mathrm{local}\text{-}y = \frac{t_{\rm{up}} - t_{\rm{down}}}{2} \cdot v_{\rm{sig}}
    \end{equation}
    with $t_{\rm{up}}$ as the digi's leading edge time at the upper channel (read out point at local-$y = +13.5$\,cm ), $t_{\rm{down}}$ the digi's leading edge time at the lower channel (read out point at local-$y = -13.5$\,cm) and $v_{\rm{sig}}$, the signal propagation speed along the strip. $v_{sig}$ can be determined by fitting a box function to the reconstructed local-y distribution with the physical length of the strip, 27\,cm, as input and the signal velocity as a free parameter.
    For all USTC counters, a signal velocity of $v_{\rm{sig}} = 18.23$\,cm/ns and for all THU counters, $v_{\rm{sig}} = 16.45$\,cm/ns was used. The difference in $v_{\rm{sig}}$ arises from the differences in the MRPC design (see Tab.~\ref{tab:tech_detais}).

    If a GET4 on one side of a strip has to resynchronize with the rest of the system (called ``GET4 dropout'' in the following) or is for other reasons inefficient while a signal travels along the strip, the other side may still detect the signal. About 95\% of all GET4 dropouts are only ``single-sided''. 
    Instead of extracting the hit position from the time difference of two digis, we can use the location of an intersection on the counter surface from a TPC track to determine the local-$y$ position of the hit. Since we know the signal velocity, we can subtract the travel time along the strip from the recorded digi time to retrieve the original hit detection time. This single-sided hit reconstruction comes at the cost of reduced time resolution by a factor of $\sqrt{2}$, but recovers hits that would be otherwise lost.
    Depending on the data set, up to 10\% of all eTOF hits are single-sided. In rare cases, all GET4s on one side of a counter can dropout for a full run. To avoid temporary holes in the acceptance, those single-sided hits are reconstructed as well.
    
    Since a single particle can induce a signal on multiple strips, hits close in local-$y$ on neighboring strips are merged.    
    The position and time of the merged hit are the averages of the contributing strips. The cluster-size is defined as the number of strips contributing to a merged hit. The average cluster-size has a typical value of 1.3 hits for fully operational MRPCs.
    Single-sided hits are not merged as their local-$y$ position is unknown prior to track matching.

    Once the hit merging is done, the TPC tracks are extrapolated to the three eTOF planes and their intersection points on the individual counter surfaces are determined. 
    If multiple hits are close enough to a track to form a match, the closest hit is matched. Double-sided hits are given priority over single-sided ones.  The maximum matching distance is chosen to be 7\,cm in local-$x$ and 10\,cm in local-$y$.

\begin{figure*}[!h]
    \centering
    \includegraphics[width=0.9\textwidth]{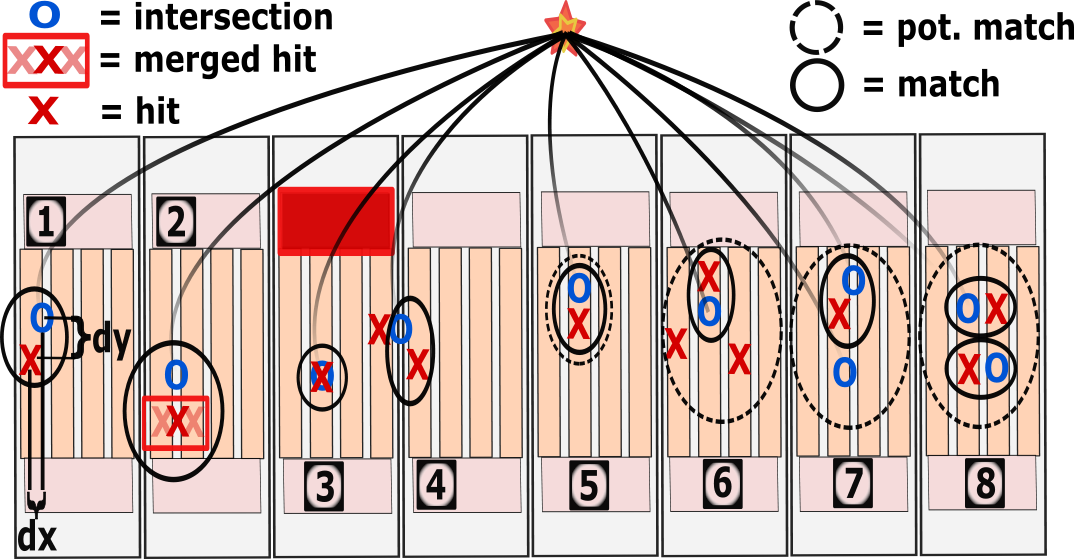}
    \caption{Overview of match cases. 
    (1) hit matched to a track intersection with a distance of d$x$ in local $x$ and d$y$ in local $y$ coordinates;
    (2) hits on neighboring strips merged into one before matching with intersection;
    (3) single-sided hit matched to intersection on strip with ``bad'' Get4;
    (4) double-sided hit matched with priority over single-sided one;
    (5) single hit matched to single track;
    (6) one out of multiple hits matched to track;
    (7) single hit matched to one out of multiple tracks;
    (9) cluster of multiple hits and multiple tracks matched to one another. 
    }

    \label{fig:hitmatch}
\end{figure*}

    In more central events with higher multiplicity, the fraction of match cases where multiple hits and tracks are in competition with each other rises and increases the likelihood of matching the wrong hit-track pair. 
    As a consequence, matches where multiple match combinations are possible have a lower signal-to-noise ratio than those without ambiguities. 
    To enable analyzers to tailor the data sample to the needs of their analysis, the information on the match type (single-sided hit or double-sided hit) and match case (single hit to single track, multiple hits to single track, single hit to multiple tracks, multiple hits to multiple tracks), as well an indication that the hit was located in the overlap region of two counters is stored in a match-flag. Fig.~\ref{fig:hitmatch} sketches the possible match flags.

\subsection{Spatial alignment}
\label{cal}

	To enable proper track matching, the hit positions have to be calibrated to account for varying cable lengths and other propagation delays. Each channel is assigned a linear offset. The offsets are determined by the distributions of the differences (accessed via local-$y$) and the sums (accessed via the time of flight) of the digi times forming hits. 

	For a first, coarse-grained position calibration, the mean of all hits on a strip is shifted to zero since the distribution along the strips (local-$y$) is expected to be uniform. To improve the accuracy further, a box function is fitted to the hit distribution allowing the alignment of the strips with respect to each other with a precision of about 1\,mm.
	The global positions of the counters are then aligned with the TPC track intersections to minimize geometrical discrepancies.
    The distribution of the distance from hits to track intersections in local coordinates ($\Delta x$ and $\Delta y$) is fitted with a Gaussian and shifted to zero. Typical values are on the order of 0.5\,cm in local-$y$ and 1\,cm in local-$x$.
    
    \subsection{Time calibration}
	The phase relationship between the start time provided by the bTOF electronics and the arrival or stop time provided by eTOF electronics must be learned from the data. The phase offset is determined using all the events from each run.
	On every strip, the expected time of flight is calculated for all hits matched to tracks within a momentum range of about 0.5 to 3.0 GeV/$c$, assuming all particles have the mass of a pion.
	The calibration offset can then be determined by a Gaussian fit of the distribution generated by subtracting the expected time of flight from the measured time of flight. 
	The precise momentum range within the interval given above is chosen for each collision energy such that the contamination from kaons and protons is minimized. Calibrated offsets are introduced for every counter. 

    The preamplifier PADI-X discriminates at a constant threshold. Therefore, the resulting arrival time depends on the rise time of the signal.
    The rise-time dependence is directly correlated with the time over threshold (ToT). A time-walk or slewing correction based on ToT removes the rise-time dependence.
	This calibration has to be performed for each channel separately as the individual preamplifier gains are not equal.  
	The expected time of flight for pions is subtracted from the measured time of flight in 25 ToT bins. 
    A Gaussian fit of the resulting distributions for each of the 25 ToT bins determines the calibration offset for each bin and channel.

    \subsection{Error handling}

	The GET4 ASICs are designed to resynchronize once a mismatch of the internal clock counter and the external sync signal is detected.
	If the sync signal arrives too close to a clock cycle edge, the jitter of the two signals can cause the sync signal to be assigned to the wrong clock cycle and shift the GET4 time relative to its ``true'' value by 6.25\,ns (called ``clock jump'' in the following).
	This will cause all time information processed by a given GET4 to be shifted accordingly, causing all hits built with a digi from that GET4 to be either 3.125 ns too early or too late and shifted by about 50\,cm in local-$y$.

    \begin{figure*}[ht!b]
        \centering
        \includegraphics[width=0.9\textwidth]{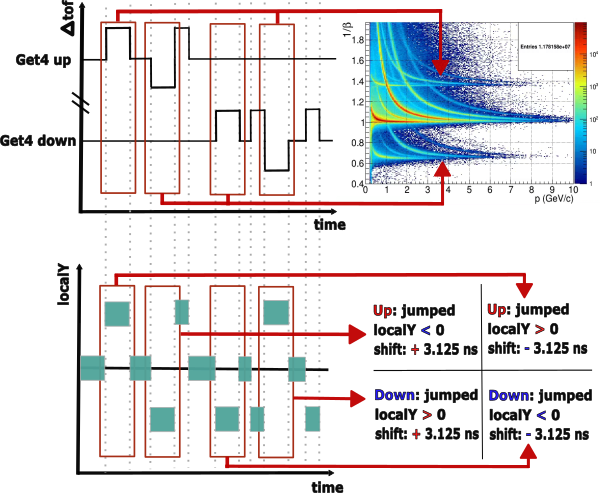}
       \caption{Sketch of the possible states of a GET4 pair.
       Top left: digi time minus expected digi time vs. time of recording on upper and lower GET4.
       Bottom left: local $y$ positions corresponding to the states of upper and lower GET4.
       Top right: 1/$\beta$ vs. momentum showing sidebands shifted by 6.25 ns caused by jumps.
       Bottom right: potential states of GET4 pair if jumping side is known from time spectrum.
    }  
    \label{fig:jumpsketch}
    \end{figure*}
	
	Whether a hit is affected by a clock jump can be determined by the shift in its local-$y$ position.
	The maximum travel time for a signal across the full length of the strip is on the order of 2\,ns. The signal propagation speed along strip is about 18\,cm/ns, thus the proper local-$y$ positions do not overlap with the positions of jumped hits.
    
	As there are two GET4s (top and bottom) and two possible jump directions (early or late), the state (top good - bottom early, top good - bottom late etc.) of a GET4 pair cannot be determined from a single jumped hit constructed from only 2 digis (see Fig.~ \ref{fig:jumpsketch}).

    To resolve the ambiguity we first determine which side of a GET4 pair jumped.
    This is done by comparing the measured digi times to the expected digi times, assuming all particles travel in straight lines at the speed of light. The time offset due to the time a signal needs to travel from the hit position to the read out points is taken into account as well. 
	By averaging the difference between the measured and the expected digi times over multiple consecutively jumped hits, the jumping side can be identified for the vast majority of scenarios. 
	Wrong identifications arise in significant numbers only if less than $\sim$ 5 consecutively jumped hits are available or state changes occur so fast that each state to be determined has to be estimated from fewer than $\sim$ 5 hits.
	Another difficult case occurs if the hit sample for averaging contains a high fraction of hits arriving later than the pion time of flight plus 6.25\,ns. In those cases, an actual late hit cannot be distinguished from an early hit that jumped to a later time.
	Once the jumping GET4 is identified, the direction of the jump in local-$y$ fully determines the state of the GET4 pair and the hit time can be shifted accordingly.
	Since the method requires knowledge about multiple 
    time-ordered events distributed over several files, the GET4 states cannot be determined at production time.
    As a consequence, this state information has to be treated like a calibration parameter in order to be compatible with STAR's production workflow. 	
	This information has to be stored and retrieved in the calibration data base during production.
    Due to memory limitations of the database, not every state for every GET4 and event can be saved.
    Even the state changes are too numerous to save.
	Instead, the fact that individual GET4 pairs tend to have a preferred jump state which accounts for the majority of jumped hits is exploited.
	For each run and GET4, either a default ``jump-side'' or ``jump-direction'' is chosen which in combination with the shift in local-$y$ determines the correct GET4 state. Only those states which explicitly disagree with their default state are stored in the database.

\section{eTOF Performance}
\subsection{Geometric acceptance}
\begin{figure*}[ht]
    \centering
    \includegraphics[width=\textwidth]{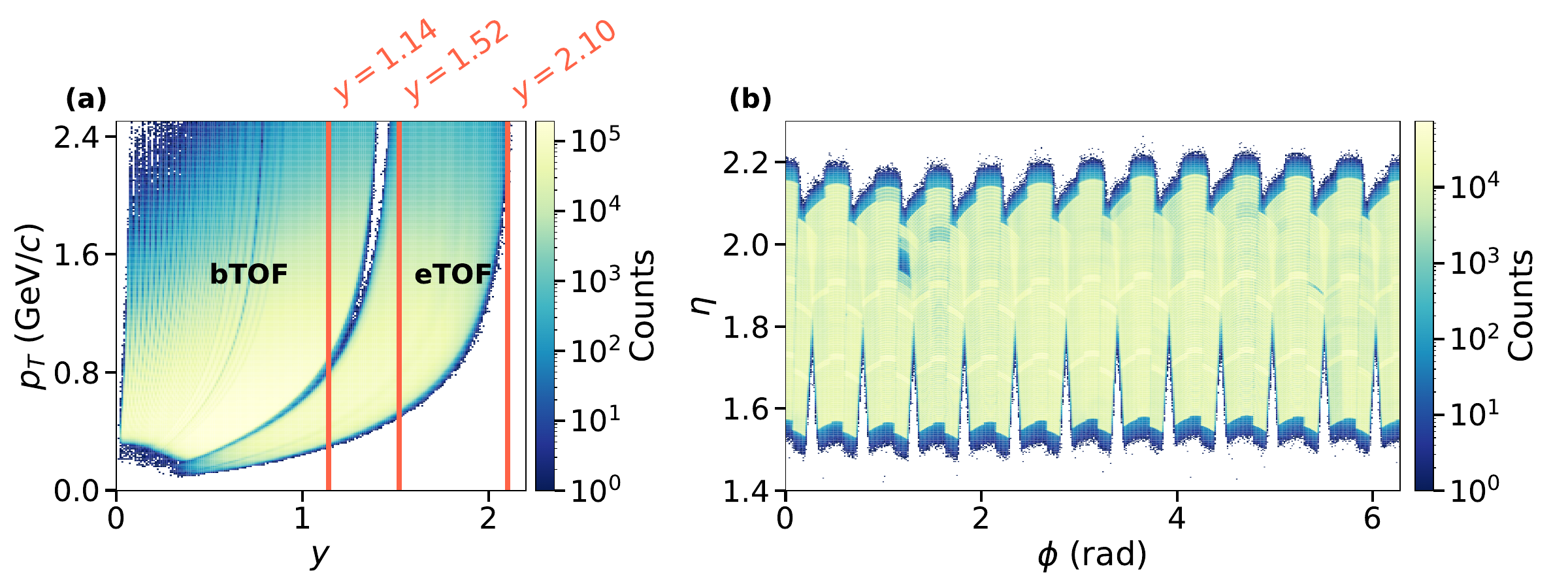}
    \caption{STAR eTOF measurements in Au+Au collisions at $\sqrt{s_{_{NN}}} = 4.5$ GeV: (a) STAR TOF acceptance in proton rapidity $y$ and $p_T$. Also shown in red is midrapidity, $y_{cm}=0$, for a number of center-of-mass energies (3.2, 4.5 and 7.7 GeV from left to right). (b) eTOF acceptance in pseudorapidity $\eta$ and azimuthal angle $\phi$.  }
    \label{fig:acceptance}
\end{figure*}

\begin{figure*}[ht!b]
    \centering
    \includegraphics[width=\textwidth]{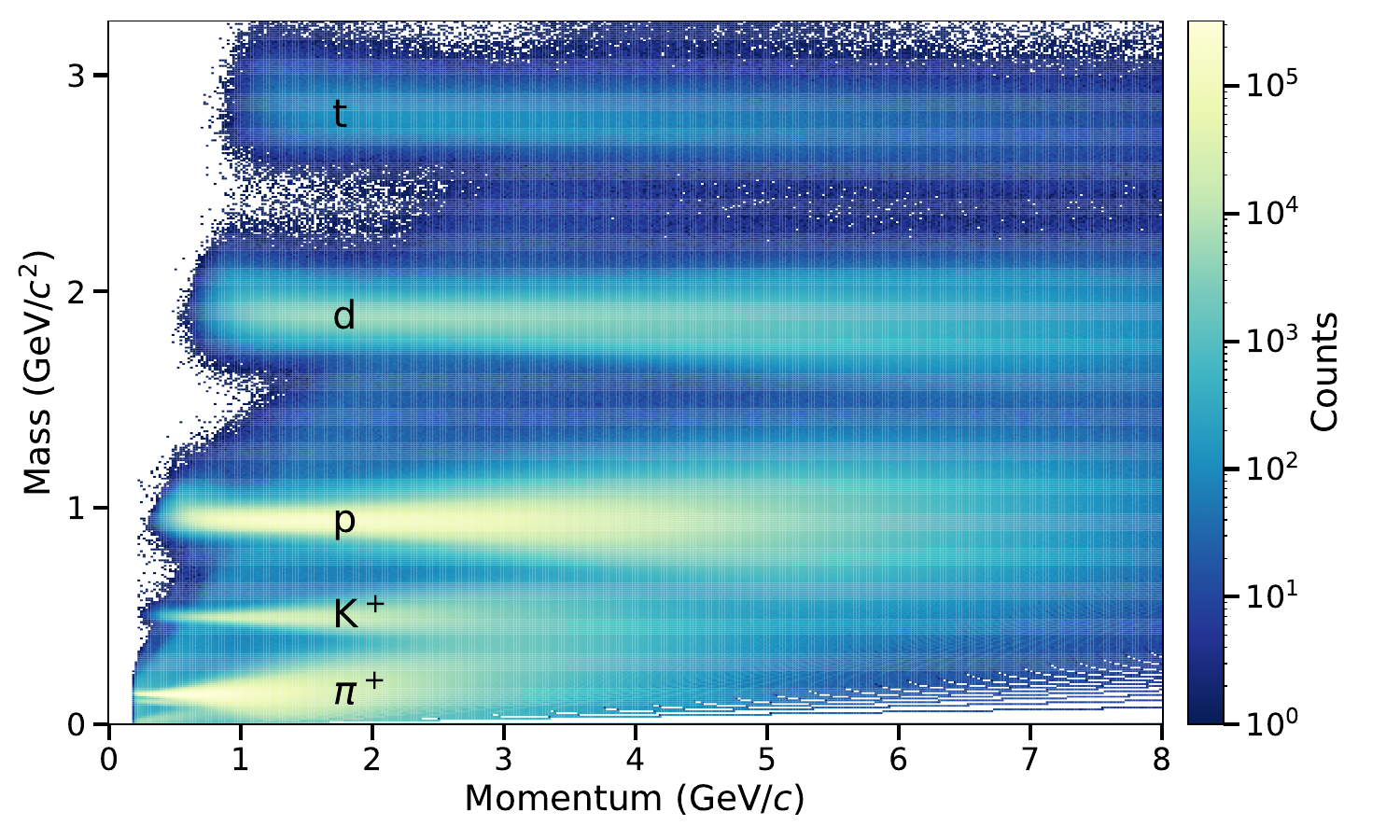}
    \caption{Mass distribution as a function of momentum of positive particles measured by eTOF}
    \label{fig:mass}
    \end{figure*}

To orient the reader, Fig.~\ref{fig:acceptance}(a) shows the measured acceptance of the STAR TOF systems as a function of transverse momentum, $p_T$, and rapidity of protons ($y$). A gap can be seen between eTOF and bTOF which is the result of a physical gap between the two systems, as illustrated in ~Fig.~\ref{fig:etof_yz}. $y_{cm}$ denotes the center-of-mass rapidity in the laboratory frame. $y_{cm}$ increases with beam energy to higher $\eta$ regions of the detector. It is seen that midrapidity at 4.5 GeV and greater is only covered by eTOF, making eTOF critical for physics analysis at these energies. Figure~\ref{fig:acceptance}(b) depicts the eTOF acceptance in pseudorapidity vs. azimuthal angle in FXT geometry.
The inclusion of eTOF also extends the pseudorapidity coverage of the central barrel in the collider configuration, extending the particle identification from $\eta < 1.5$ to $\eta < 2.17$ in FXT mode. Fig.~\ref{fig:mass} shows the mass distribution extracted by eTOF for positively charged pions, kaons, protons, deuterons, and tritons. Particle mass peaks are well separated at low momentum but begin to merge at higher momentum due to the finite time resolution and saturation of the particle velocity.

\subsection{Time resolution}

    The time resolution of the eTOF system determines its reach in PID.
	The system time resolution has contributions from the eTOF detector itself (MRPC and FEE resolutions) and the start time. 
	Momentum and path length uncertainties add only a minor contribution of about 5 ps which is incorporated in the eTOF resolution. 
	The combined resolution of the MRPC and frontend electronics can be estimated from particles in the overlap region of eTOF that have two time measurements.
	We consider particles with a trajectory close to a straight line originating from the primary vertex with a momentum larger than 1\,GeV/$c$ and a velocity of 0.98 $< \beta <$ 1.02.
    In addition, only particles with an energy loss in the TPC close to the value expected for pions (deviation less than 3 $\sigma$) are evaluated.    
    The time resolution of the MRPC counters can be obtained from matched hit pairs by projecting the measured time of the hit in the back counter (REF) to the front counter (DUT) employing the velocity derived from the REF counter.   
    When fitting the distributions of all hit pairs with a Gaussian, an average time resolution of 93\,ps is obtained.

     \begin{figure*}[ht!b]
    \centering
    \includegraphics[width=0.95\textwidth]{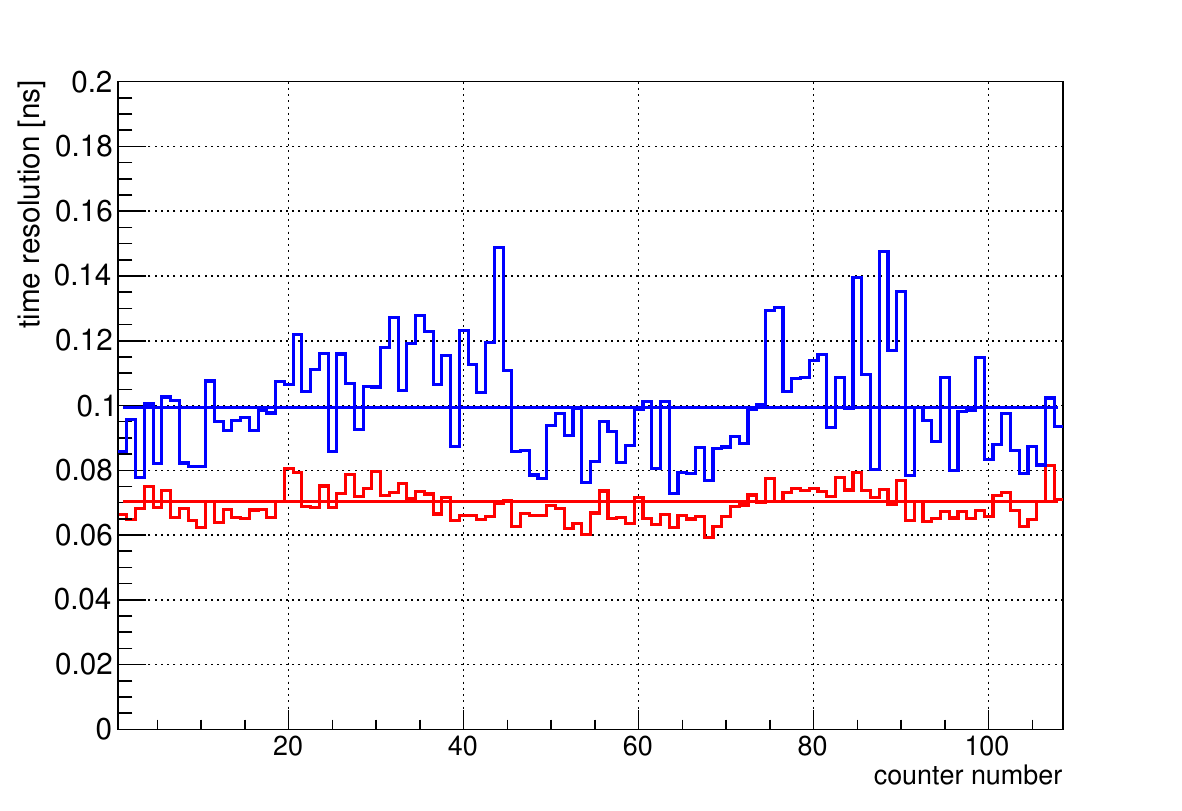}
    \caption{Time resolution of all 108 counters for double-sided matches (red), on average 71.4 ps and single-sided hits (blue), on average 99.3 ps.}
    \label{fig:res2}
    \end{figure*}  
	Assuming that REF and DUT contribute equally to the combined resolution, this results in an average time resolution of 66\,ps per counter.
	This result is in good agreement with the average system time resolution of 71.4 ps (69.1 ps for USTC and 76.0 ps for THU counters) obtained by comparing measured and expected times of flight for pions as described in Sect.
    \ref{cal}, factoring in a start time resolution of about 25 ps. Table~\ref{tab:resolution_contributions} shows the contributions to the detector resolution from various sources.
\begin{table}[h]
\centering
\renewcommand{\arraystretch}{1.2}
\begin{tabular}{l@{\hspace{3cm}}l}
\hline
Source & Contribution \\
\hline
bTOF $t_{\mathrm{start}}$ & $\sim 21$~ps \\
MRPC & $\sim 60$~ps \\
GET4 & $\sim 20$~ps \\
PADI-X & $\sim 15$~ps \\
TPC tracks & $< 5$~ps \\
\hline
\end{tabular}
\caption{Known contributions to the eTOF system time resolution.}
\label{tab:resolution_contributions}
\end{table}
	The system resolution for all 108 counters is shown in Fig.~\ref{fig:res2}. 
	The resolution of the USTC MRPC's is slightly better than that of the THU counters that were designed with a stronger focus on rate capability.
	The variations between the counters have contributions from production tolerances of the MRPC's themselves, the position in the eTOF detector, imperfections in the calibration, and variation in clock distribution stability.

\begin{figure*}[ht!b]   
    \centering
    \begin{minipage}{0.49\textwidth}
    \centering
    \includegraphics[width=\columnwidth]{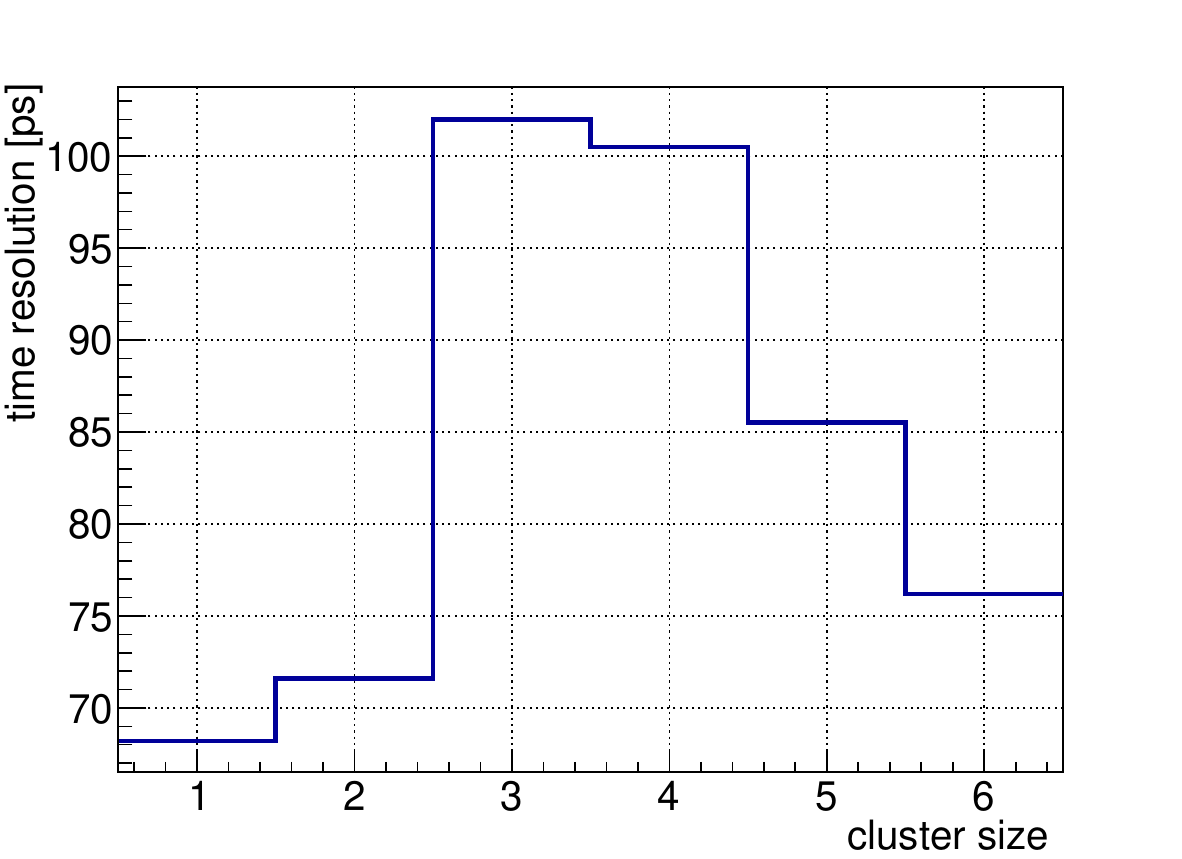}
    \caption{Time resolution dependence on the cluster size.  }
    \label{fig:resclsz}
    \end{minipage}\hfill
    \begin{minipage}{0.49\textwidth}
    \centering
    \includegraphics[width=\columnwidth]{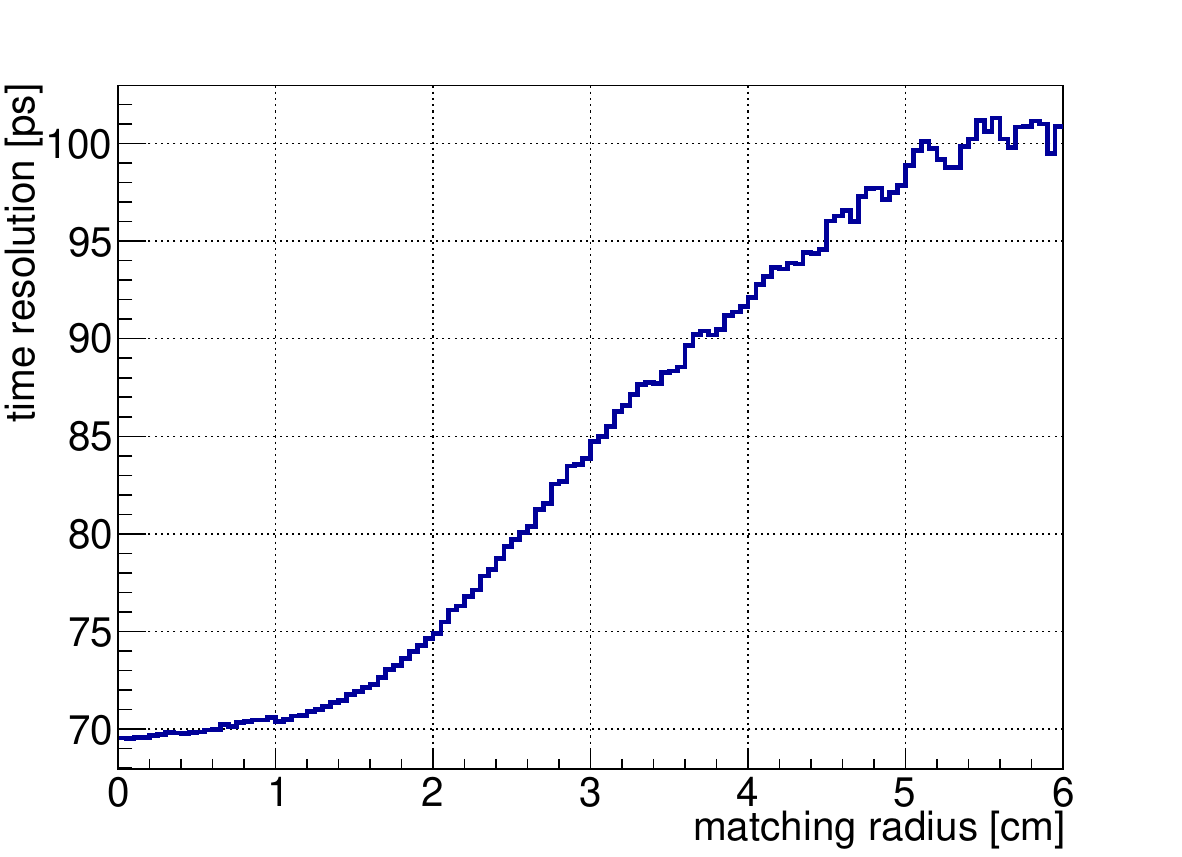}
    \caption{Time resolution dependence on matching distance. } 
    \label{fig:resrad}
    \end{minipage}  
    \end{figure*}

	Significant differences in terms of resolution are seen comparing single-sided and double-sided matches.
    The match case, on the other hand, has only an indirect and minor impact on the resolution as it mainly affects the signal-to-background ratio, not the timing of properly matched hits. The resolution for the most complex case (multi hit - multi track match) is on average just below 85\,ps while the simpler match cases show a resolution of about 70\,ps. 
    For hits with a cluster size greater than one, one might expect an improvement of the resolution. But imperfections in the walk calibration and the differences in electronics resolution in between strips outweigh the benefit of multiple, independent measurements (see Fig.~\ref{fig:resclsz}). The vast majority of hits (98.9\%) have a cluster size of one or two, thus the system resolution is only slightly impacted by hits of size three or more.

    The time resolution is more dependent on the distance of the hit to the matched track (see Fig.~\ref{fig:resrad}). The track-match distance is one of the main leverage points for tuning the tradeoff between efficiency and purity. Up to a matching distance of about 3 cm which includes 90\% of all matches, the system resolution remains better than 80 ps. 
	
\subsection{Efficiency and Purity } 
The eTOF efficiency is calculable by considering the number of TPC tracks that project towards the eTOF acceptance and the eTOF hits that are matched to a TPC track. To quantify the efficiency, helical TPC tracks are extrapolated into the eTOF acceptance region. An eTOF hit is considered to match a TPC track if its position lies within a specified tolerance in the $x$ and $y$ directions relative to the track's projected position in the $xy$ plane at the $z$-coordinate of the eTOF hit. The matching efficiency can then be determined as the ratio between the number of matched eTOF hits relative to the total number of TPC tracks pointing towards the eTOF acceptance. The $xy$ tolerance can be adjusted depending on the user's needs. A tighter requirement will allow for a larger fraction of high-quality eTOF track matches, improving the time resolution but lowering the efficiency. Wider $xy$ tolerances will give a higher background, but also improve the matching efficiency.

\begin{figure*}[ht!b]
    \centering
    \includegraphics[width=\textwidth]{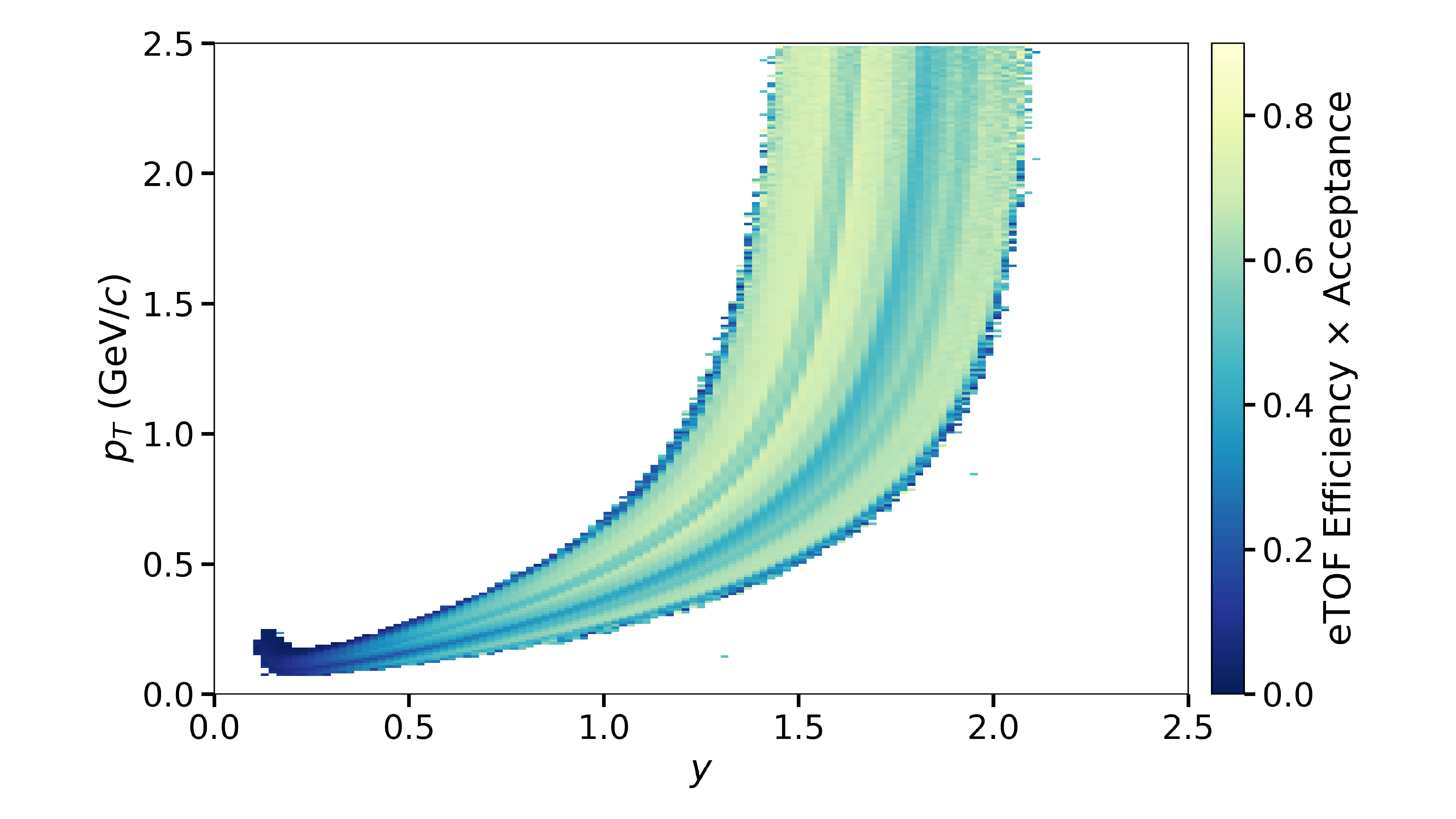}
    \caption{Proton matching efficiency as function of rapidity and transverse momentum.}
    \label{fig:MEyvspt}
    \end{figure*}

Particle-species-dependent matching efficiencies can also be determined using this technique by utilizing the TPC $dE/dx$ measurement and comparing it to the Bichsel model \cite{RevModPhys.60.663, bichsel2006method}. The Bichsel model accounts for shell effects and energy-loss fluctuations providing improved resolution over the Bethe-Bloch formula, particularly in the low to intermediate momentum range. Bichsel curves are generated for different species, depending on the 
mass-to-charge ratio. Fig.~\ref{fig:MEyvspt} Shows the matching efficiency as a function of rapidity and transverse momentum for protons. The shape of the distribution is dictated by the pseudorapidity acceptance of the eTOF wheel.

$dE/dx$ distributions at various rapidities and transverse momenta are fitted with Gaussians for each particle species and the central values and standard deviations ($\sigma$) are extracted. The $dE/dx$ of particles of a given momentum as measured by the TPC are then compared to the $dE/dx$ predicted by the Bichsel curves at that momentum. If the $dE/dx$ value measured by the TPC falls within 2$\sigma$ from the predicted Bichsel value, the track is accepted as a well-identified species. For example, a proton matching efficiency can be calculated from the ratio of the total number of eTOF hits that match to a well-identified proton track from the TPC to the total number of proton tracks. However, at certain momenta, the bands of the different Gaussians will merge making the extraction of a reliable standard deviation more difficult. In these instances, care must be taken when extracting a matching efficiency using the method described above. For instance, protons of momenta greater than 1.35 GeV suffer contamination from pions in $dE/dx$.

\begin{figure*}[ht!b]
    \centering
    \includegraphics[width=\textwidth]{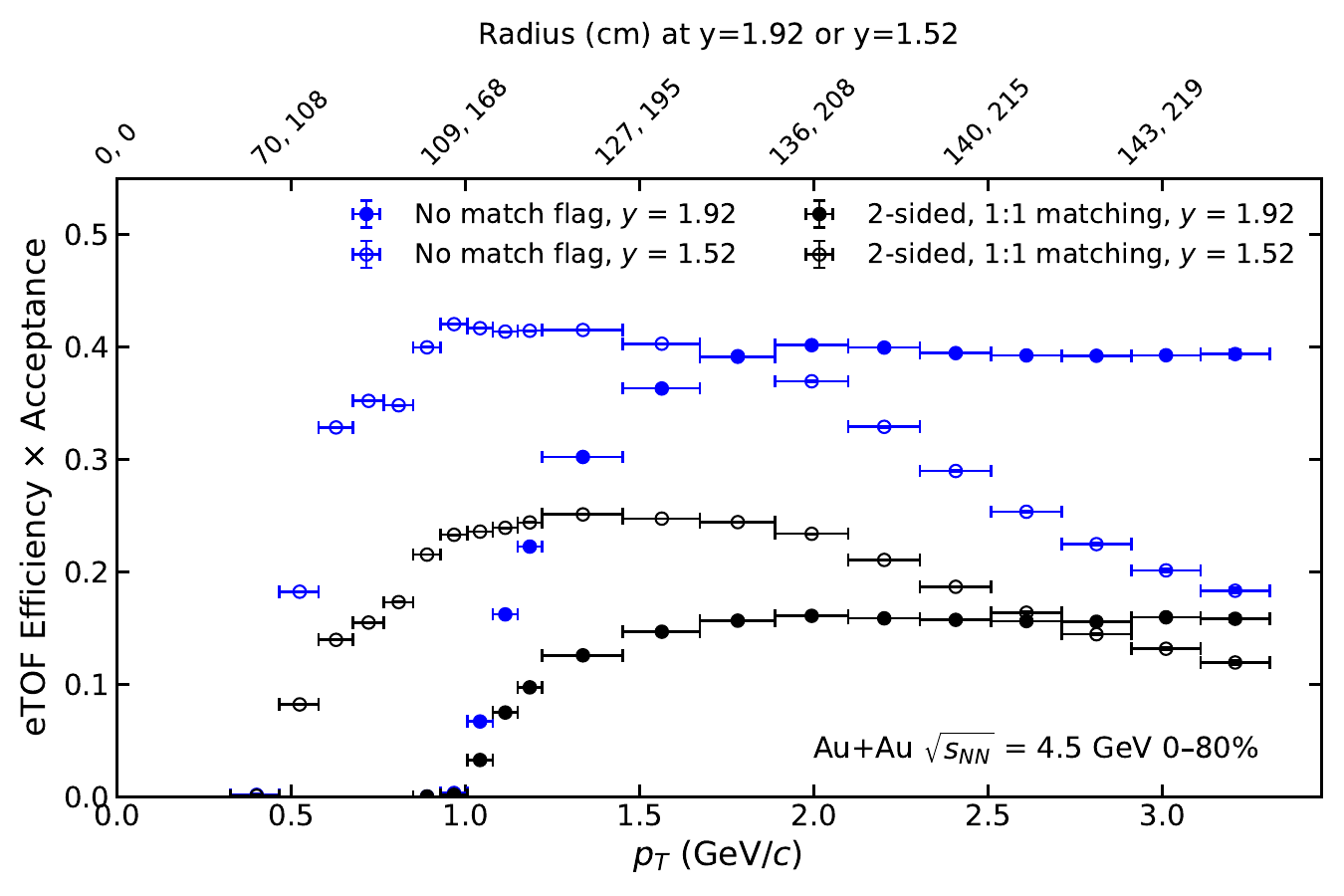}
    \caption{eTOF efficiency for protons as a function of transverse momentum. Also shown on the top axis is the radius of the track at the point of eTOF intersection, measured from $r$ = 0. Blue points show the case where there is no match flag requirement, and black points show the 2-sided, 1:1 matching case. It can be seen that including a match flag requirement lowers the matching efficiency (open markers). Efficiency curves are shown for two rapidities ($y = 1.52$ and $ y = 1.92$). The top axis shows the radius of tracks at the point of eTOF intersection for $y = 1.52$ and $y = 1.92$.}
    \label{fig:efficiency}
\end{figure*}

Figure~\ref{fig:efficiency} shows the matching efficiency for protons for two rapidities as a function of transverse momentum and for different match-flag cases. Using a 2-sided hit requirement and 1:1 particle matching with the TPC decreases the efficiency by a factor of two, but improves the mass resolution. At low transverse momentum, the efficiency grows with $p_T$, because of poor constraints on the TPC helix. Low-$p_T$ particles also are more affected by material between the eTOF and the TPC endcap causing a drop in efficiency.  At larger $p_T$, the efficiency decreases due to the geometry of eTOF. The counters cover less of the total area and physical gaps in the acceptance exist. Tracks at rapidity $y = 1.92$ (closed markers) never reach the radii at which the acceptance diminishes and a plateau is observed. This contrasts to the tracks at $y = 1.52$ (open markers) where the outer reaches of the eTOF wheel are accessed causing a decrease in eTOF efficiency as a function of transverse momentum. 

\begin{figure*}[ht!b]

    \centering
   \includegraphics[width=0.95\textwidth]{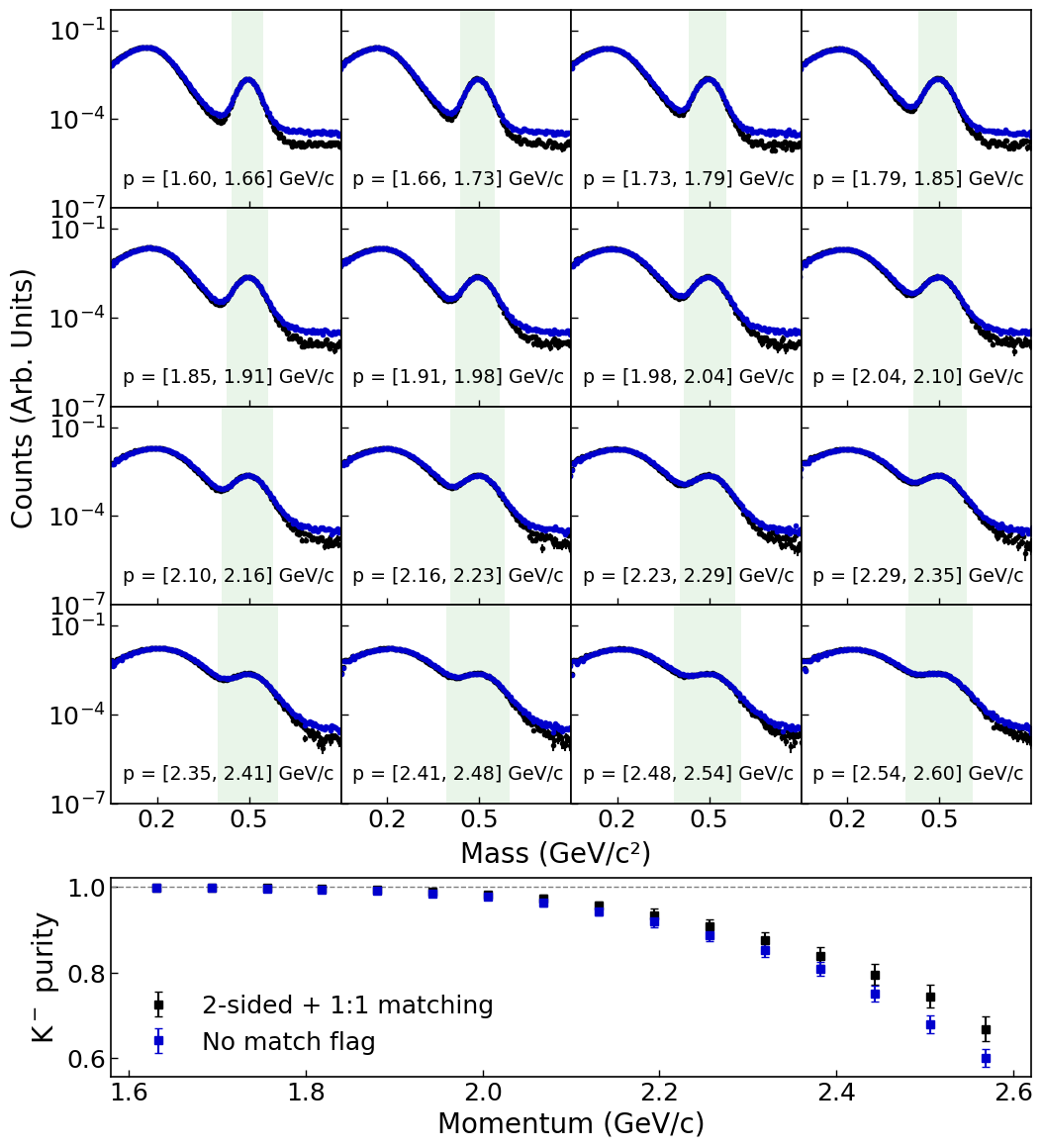}
    \caption{Upper panel: eTOF measurements of mass distributions for negative particles measured by eTOF in different momentum bins for different match-flag cases for Au+Au collisions at $\sqrt{s_{NN}} = 4.5$ GeV. Antiprotons and heavier antiparticles are heavily suppressed at this beam energy and are not shown. Green shaded boxes show the $\pm 2 \sigma$ range around the centroid of the mean value of the peak extracted by fitting the peaks to a student t distribution. Lower panel: K$^-$ purity within 2$\sigma$ for the different match cases. Using the match-flag, the background is reduced by roughly a factor of five. }
    \label{fig:flagComparison}
\end{figure*}

Figure~\ref{fig:flagComparison} compares the purity of K$^-$ particles with and without using the match flag as a function of momentum. The background from pions is determined by fitting the $\pi^-$ peak, and extrapolating to the $\rm K^-$ region. K$^-$ are produced less frequently than K$^+$ at these energies, making them harder to separate from pions. At higher momenta, the match flag allows for a higher K$^-$ purity. Such cuts allow for separation of charged particles to higher momenta, extending the available momentum that is available for study.

    \subsection{1/$(\beta\gamma)^2$ cuts} 
\begin{figure*}[ht!b]
    \centering
    \includegraphics[width=\textwidth]{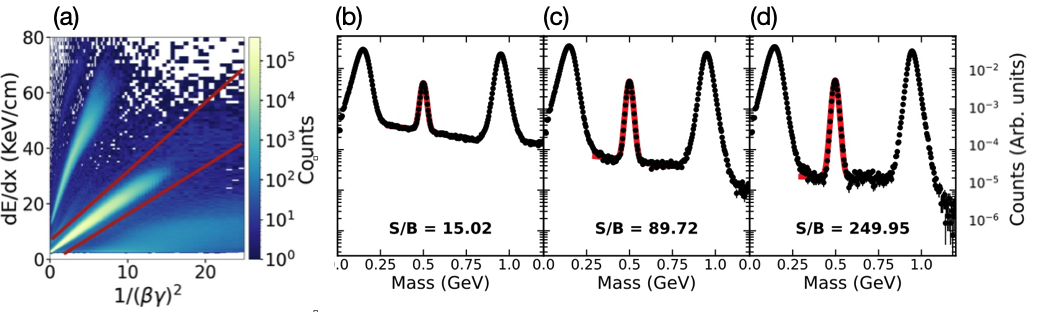}
    \caption{(a): $dE/dx$ as a function of $(\beta \gamma)^{-2}$. Red lines indicate cuts that are implemented to extract the signal in (b - d). Above  the red lines are bands associated with charge two and charge three tracks. Below the red line is background from erroneous matches with tracks containing poor timing information. (b - d): Mass distributions for positively charged particles at $p = 1$ GeV/$c$ extracted by eTOF using different quality requirements. Also indicated in each panel is the ratio between the kaon signal counts at the highest point ($S$) and the value of the background function at that point ($B$), which is taken to be a linear function. Red lines indicate fits to a student-t distribution. (b): No requirements. (c): Requiring $\Delta x$ and $\Delta y <$  1 cm and $(\beta \gamma)^{-2}$ cuts. (d): The same cuts as in (c) with the addition of a double sided, 1:1 Get4 hit matching requirement. }
    \label{fig:bgCutsAndMass}
\end{figure*}

It is possible to show from the relativistic Bethe-Bloch equation that the energy-loss of particles of the same charge will fall on a common line when plotted as a function of $1/(\beta\gamma)^2$ \cite{bethe2014braking}. Using this fact, it is possible to reduce the background of the eTOF mass signal by plotting the $dE/dx$ measurement from the track and the $1/(\beta\gamma)^2$ measurement of the corresponding matched eTOF hit. If a given eTOF hit exhibits a $1/(\beta\gamma)^2$ measurement far from the common line for particles of a certain charge, then it can be concluded that the time of flight calculation was spurious for that hit. By requiring that eTOF hits only fall on the line for the charge of interest, it is possible to further reduce the background in the eTOF mass distribution. As a consequence, $1/(\beta\gamma)^2$ is preferred as a variable to $\beta\gamma$ since low mass tracks will have large values of $\beta\gamma$ making it computationally difficult  to remove them. However if the inverse is taken, the low mass tracks are mapped close to zero and are easier to handle. Figure~\ref{fig:bgCutsAndMass}(a) shows $dE/dx$ from the TPC as  function of $1/(\beta\gamma)^2$ from eTOF. The bright band between the red lines shows Z = 1 tracks, and a similar bands can be observed above, corresponding to Z = 2 and 3. The red lines indicate a cut that can be placed to reduce the background.
    A concentration of tracks is also seen having a very small time of flight measurement and can be removed by the cut. The cut removes other background that is well away from the curve predicted by the Bethe-Bloche equation. 
    Figure~\ref{fig:bgCutsAndMass} also shows mass distributions at a momentum of 1 GeV/$c$ for three different eTOF hit quality requirements. In each case left to right the background becomes further reduced. 

\subsection{Acceptance fluctuations}
 \label{sec:acc}

	Due to the aforementioned clock jumps and GET4 drop-outs, the eTOF acceptance can fluctuate in time. 
	The clock jumps affect about 1-5\% of all hits in a given dataset and tend to be located to a large extent on certain GET4 pairs.
	The clock jump correction succeeds in $\sim$ 99.5\% of the cases for the full system. The correction succeeds in $\sim$ 95\% of the cases even for the most frequently jumping GET4 pairs (about 10\% of all pairs). The remaining 5\% of clock fluctuations are expected to be negligible for most types of physics analysis. 
	
\begin{figure*}[ht!b]
    \centering
    \includegraphics[width=\textwidth]{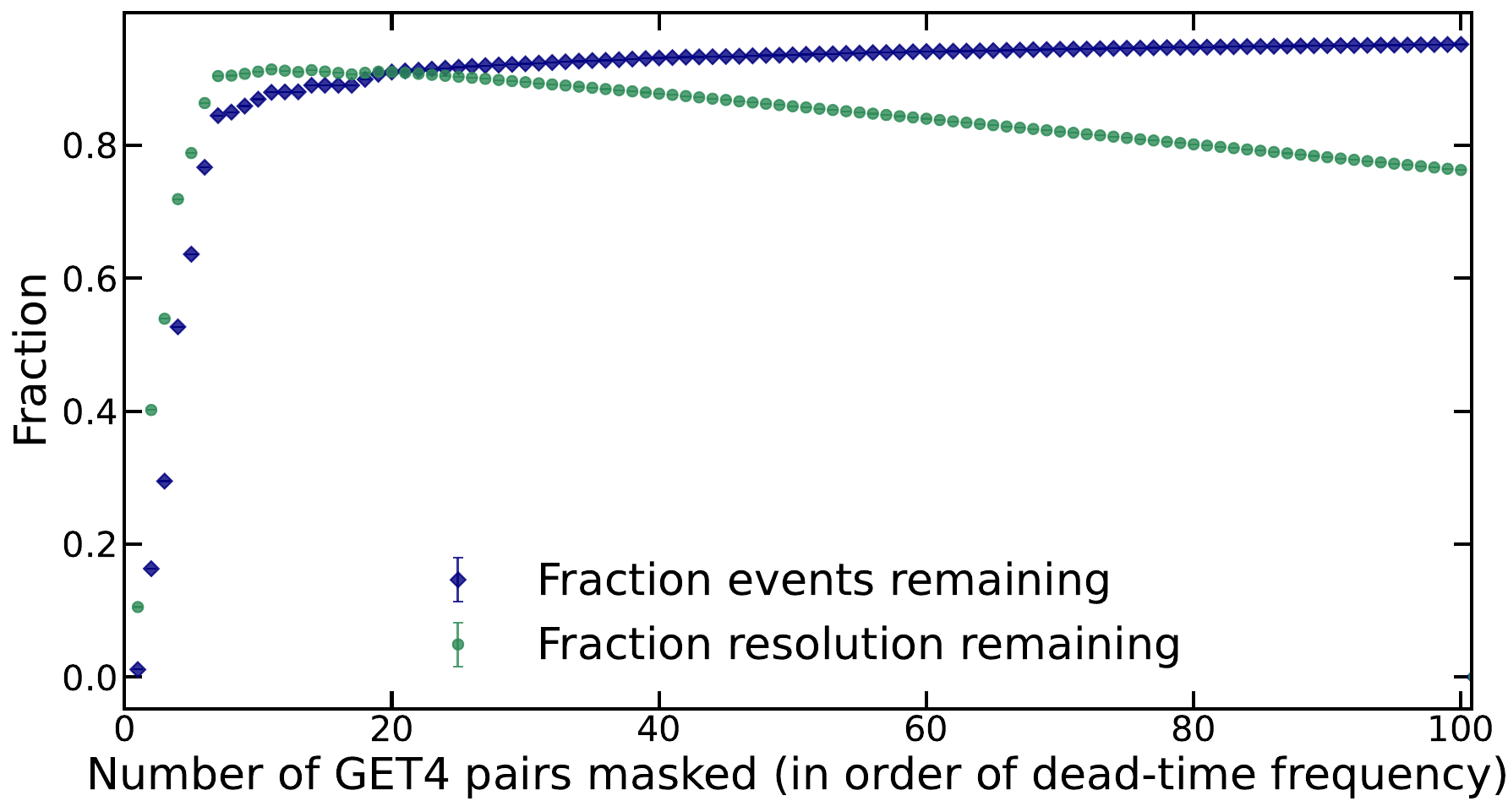}
    \caption{Blue diamonds: fraction of events remaining in the event sample after masking GET4 pairs from the active acceptance at $\sqrt{s_{NN}} = 4.5$ GeV STAR FXT. More masking leads to fewer events with a dead GET4 and a larger event sample. Green circles:  fractional proton number resolution loss caused by the GET4 masking procedure. GET4 pairs are masked in order of the frequency in which they are dead. The maximum value of this plot occurs at 11 GET4 pairs, can be taken as the optimal amount to mask at 4.5 GeV.}
    \label{fig:fracEventsKept}
\end{figure*}

	The GET4 drop-outs are common enough to cause fluctuations on an event-by-event basis and  leave almost no event entirely unaffected. For single particle observables, this effect can be mitigated by averaging over all events.
	If a fixed acceptance is crucial for a given analysis, e.g. fluctuations of conserved net-baryon number, a slightly more involved strategy can be implemented. Only a small fraction of GET4 pairs account for the majority of drop-outs. 
    The following method allows the generation of an event sample with stable acceptance.
    First, GET4 pairs are sorted by their reliability via their average number of set ``Get4Active'' flags in the full event sample. 
	Using this information, the most unstable GET4 pairs are taken out of the analysis. All events are checked for an occurrence of unmasked non-active GET4 pairs. 
    If none are found, the event is added to the event sample with stable acceptance as all remaining GET4s were active.
    If single-sided matches are included in the sample, an event has to be discarded only if both sides of a GET4 pair were inactive.
    By masking more of the eTOF acceptance, more events are included in the event sample since fewer fluctuating GET4s are present outside of the masked area. However, the event-by-event resolution on particle number diminishes with more masked GET4s due to the reduction in acceptance. These effects to the resolution can be optimized by considering the loss of event-wise statistics, which goes as $\sqrt{N_{events}}$, and the resolution of excess kurtosis of proton number after GET4 masking, which is proportional to the square of fraction of the remaining GET4 pairs. Multiplying the two quantities together allows the calculation of the optimal amount of GET4 pairs to mask for an eTOF dataset, and represents the fraction of net-proton number resolution remaining after the masking procedure. Figure~\ref{fig:fracEventsKept} shows the fraction of events remaining after masking GET4 pairs in order of the frequency in which they are inactive in the ~$\sqrt{s_{NN}} = 4.5$ ~GeV dataset. After masking the 11 most frequently dead GET4 pairs,  88\% of the total events can be recovered with a constant acceptance in every event. Measurements of all observables concerned with event-by-event effects can follow this prescription in order to parse out artifacts associated with a fluctuating acceptance.

\section{Physics implications}
eTOF provides acceptance at midrapidity for the FXT program of STAR. Some of the important physics implications are reported below.
\subsection{Critical Point search}
\begin{figure*}[ht!b]
    \centering
    \includegraphics[width=\textwidth]{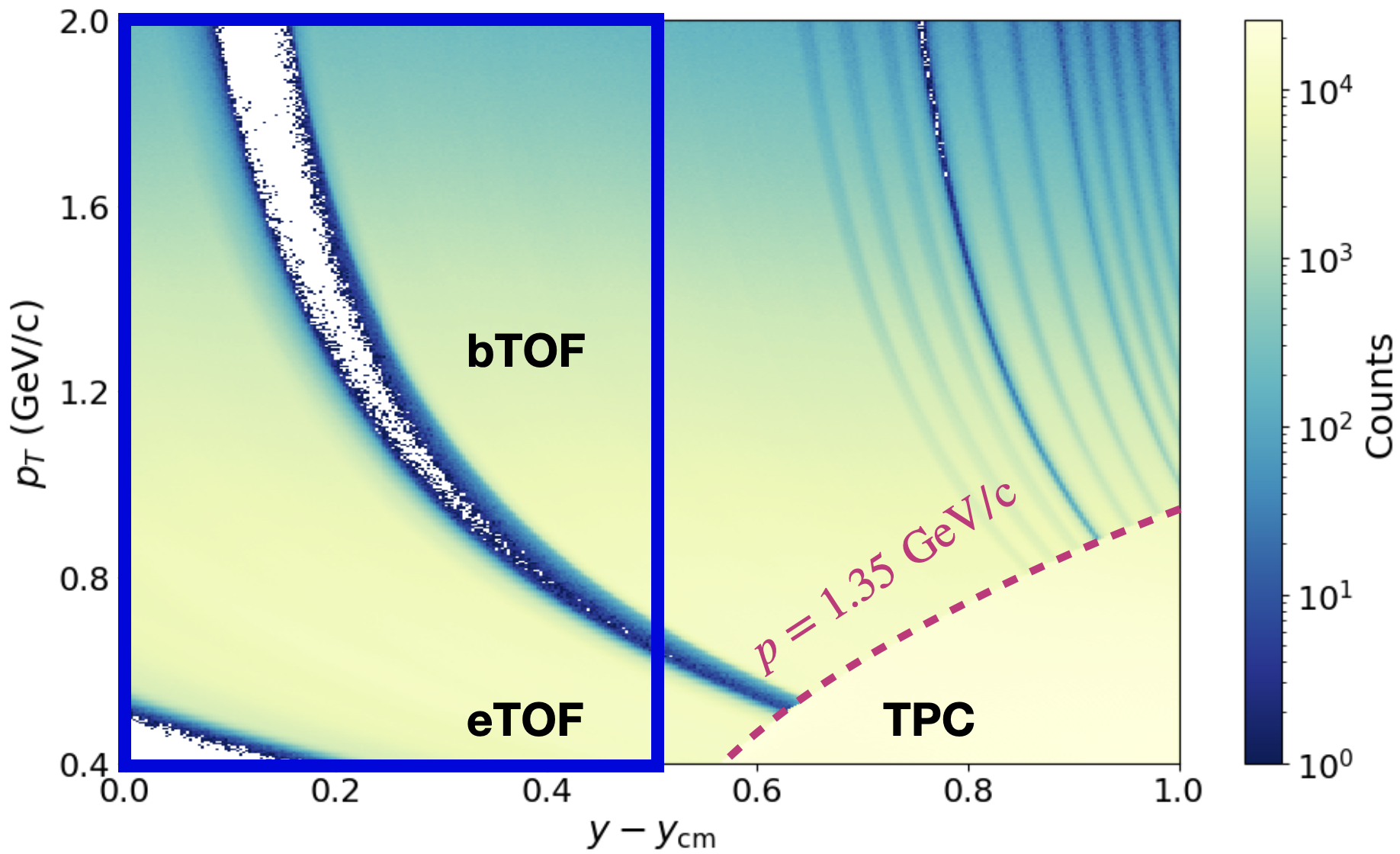}
    \caption{Acceptance and particle identification techniques used in the STAR critical point search at $\sqrt{s_{NN}} = 4.5$ GeV. The dashed line of constant momentum in the lab frame is drawn to indicate where particle contamination occurs in $dE/dx$. At $p = 1.35$ GeV/$c$, the pion band moves into the proton band, and proton purity drops below 90\% for momenta above this line. The solid blue line indicates the conventional phase space window used in the FXT net-proton number fluctuations analysis. eTOF identifies the majority of the protons measured. 
    }
    \label{fig:fluctuations_acceptance}
\end{figure*}

A key research activity at STAR is the search for a QCD critical point. The critical point would separate first and higher-order phase transitions between a hadron gas and quark-gluon plasma. The search is accomplished by measuring event-by-event fluctuations of conserved charges, such as baryon number \cite{stephanov1999event}. The net-proton number within a certain phase-space region is used as a proxy for total baryon number in the system. Fluctuations of net-proton number within the same classes of events are argued to be a signature of fluctuations associated with a critical point in the nuclear phase diagram \cite{gupta2011scale}. Higher order cumulants of the event-by-event proton number distribution are used to quantify the proton number fluctuation.  The conventional phase-space window in collider mode is from $-0.5 < y_p < 0.5$ and $0.4 < p_T < 2.0$ GeV, where $y_p = y - y_{\rm cm }$, as seen in Fig. \ref{fig:fluctuations_acceptance}. 

The asymmetric acceptance in FXT mode motivates
that the analysis is done in a half-rapidity window  ($0.0 < y_p < 0.5$). Even with a truncated rapidity window, protons still fall out of the bTOF acceptance and are then measured by eTOF, making eTOF critical for the QCD critical point studies. Figure~\ref{fig:fluctuations_acceptance} shows the analysis window used for the QCD critical fluctuations studies for the 4.5 GeV center of mass energy. As the energy increases, the analysis becomes more reliant on eTOF for proton PID. Fig.~\ref{fig:fluctuations_acceptance} also displays the two different PID techniques used in the analysis, indicated by the pink line. The TPC can be used for PID for low-momentum tracks. However, at a center of mass energy of 4.5 GeV, TOF is required within the analysis window in order to maintain particle purity. The TOF measurement is nearly hermetic within the chosen analysis window, apart from the gap in acceptance between bTOF and eTOF as well as a corner near midrapidity where tracks are forward of the eTOF acceptance. However, these gaps are controlled for by also including the same gap when comparing to model calculations. 

The fluctuation analysis at STAR varies the analysis window chosen at the lower energies in order to study systems of varying temperature and baryon chemical potential. A unique opportunity is given at $\sqrt{s_{_{NN}}} = 3.0$ GeV. Here, midrapidity is sufficiently small such that one can study fluctuations of proton number with a fully symmetric rapidity selection using eTOF. Thus it is possible to make a direct comparison to the collider analysis in an identical phase-space.

\subsection{Particle spectra}
Measuring the rapidity density distributions ($dN/dy$) of various particles is also a principle goal in BES-II. The inclusion of eTOF extends the phase-space available for study and allows for better constraints on the widths of the $dN/dy$ distribution for each particle. This can yield important physics insights \cite{E-802:1999ewk, Braun-Munzinger:1994ewq, Cassing:1999es}. For example, $dN/dy$ distributions of charged pions and kaons are known to be Gaussians that have different widths depending on the charge \cite{E-802:1999ewk}. This can be understood as a consequence of associated production, the process in which thermally produced $s$ quarks form a $\Lambda$ baryon that subsequently decays into a proton and $\pi^-$. The remaining $\overline{s}$ quark can combine with another $u$ quark to form a positively charged kaon. The result is a wider distribution of $\rm K^+$ relative to $\rm K^-$. Furthermore, the rapidity density distribution of protons allows for the study of baryon stopping, the degree to which the protons and neutrons from the nuclear targets lose rapidity during the collision. By studying the systematic trend in baryon stopping as a function of beam energy, it is possible to draw conclusions about the nature of the phase transition between a hadron gas and a QGP \cite{ivanov2010_baryon_stopping, ivanov2013_alt_scenarios, ivanov2016_epja_baryon_stopping, ivanov2015_robustness}. 

While some of these measurements have been made at the AGS and SPS \cite{ahle_excitation, na49_pik, na44_strange, e895_longitudinal_flow_2001, e802_kaon_centrality}, 
there is tension in the data between experiments that requires further experimental constraint. Moreover, charged kaon rapidity distribution measurements are still missing at the energies accessed by the STAR FXT program. eTOF enables the STAR FXT program to resolve the measured discrepancy of midrapidity kaon production, and perform the first measurement of the width of the kaon $dN/dy$ distribution at these energies.

The use of the match flag in this analysis is the opposite of the situation described for proton moments. Since hadron production measurements are on average calculated over all events, event-by-event fluctuations in detector acceptance are reflected in the efficiency calculation. This makes it unnecessary to impose strict requirements on the amount of masked GET4 readouts. Particle spectra can be improved by requiring two active GET4  measurements for a given eTOF hit. This refines the mass resolution since two independent timing measurements are used when calculating the time of flight. Single-hit to single-track matching can also be enforced by the match flag which improves particle resolution by removing ambiguity in the TPC track-to-eTOF hit matching procedure. These requirements come at the expense of eTOF matching efficiency, as shown in Fig.~\ref{fig:flagComparison}. But since most cases of particle production measurements do not require high statistics, this is acceptable. Lastly, a significant amount of background can be removed by a $1/(\beta\gamma)^2$ cut, which will further clean up the eTOF signal and give reliable and well isolated particle bands over a significant range in momentum.

\subsection{Overlap energies with collider}
\begin{figure*}[ht!b]
    \centering
    \includegraphics[width=\textwidth]{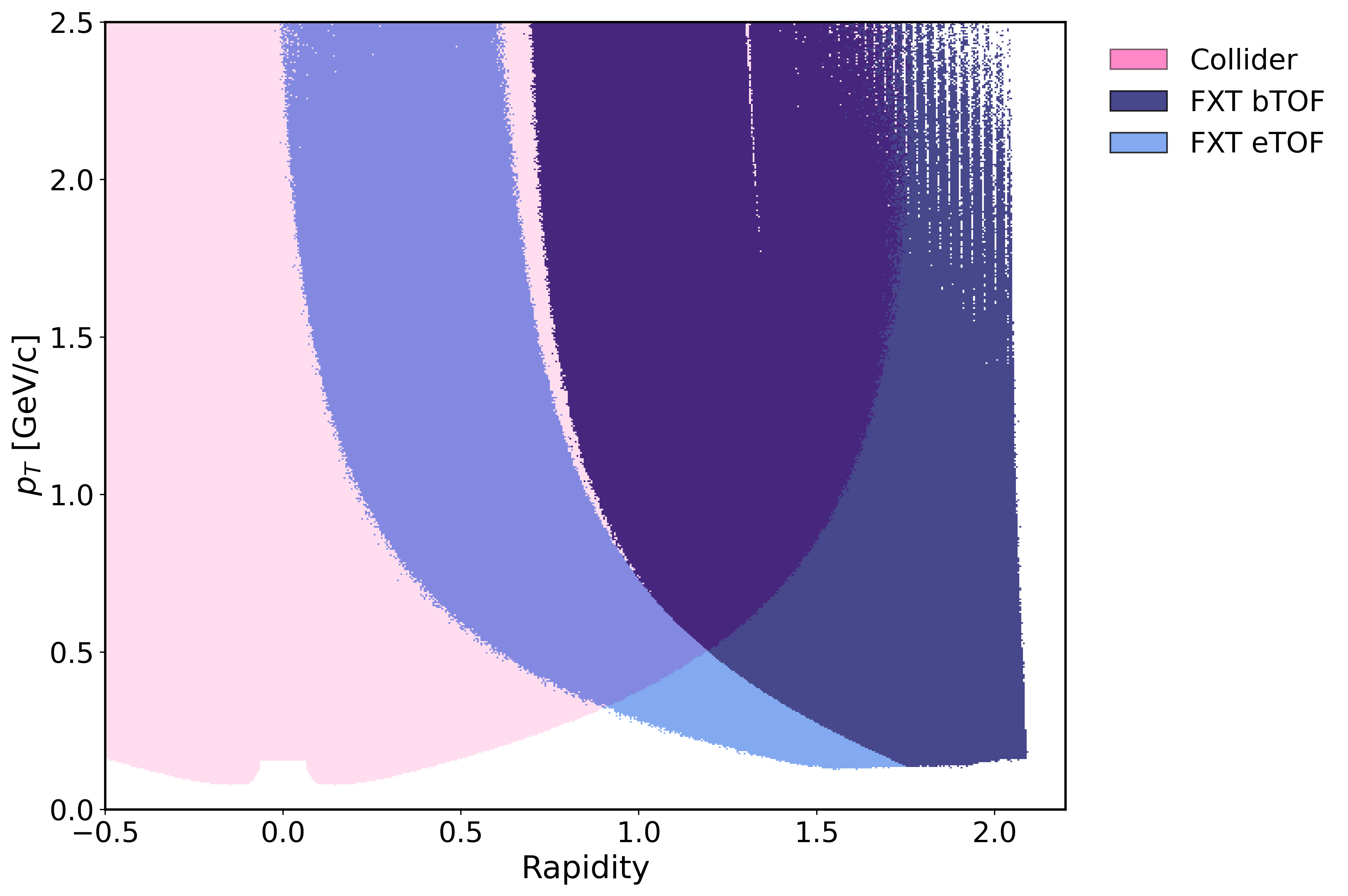}
    \caption{Acceptance overlap between collider and FXT configurations at ~$\sqrt{s_{NN}} = 7.7$ GeV. Here, extensive comparisons between the two geometries can be made in order to gain a detailed understanding of the systematic error in FXT. Collider mode is shown in pink, and bTOF is shown as dark blue, and eTOF as light blue. With the inclusion of eTOF, it is possible to make comparisons to collider mode up to nearly midrapidity.}
    \label{fig:7p7}
\end{figure*}

The FXT program includes datasets with collision energies coincident with datasets from collider mode ($\sqrt{s_{NN} } = 7.7, \ 9.2,\  11.5$ GeV). This allows for $dN/dy$ measurements of various particle species over two units of rapidity by combining the data from each configuration. With the inclusion of eTOF, there is a significant degree of overlap in phase space between the two configurations. This allows validation of the new experimental geometry of the FXT program by comparing measurements between configurations of a given observable over the same region of phase space. For example, at $\sqrt{s_{NN} } = 7.7$ GeV, there is overlap over one and a half units of rapidity where the yields of different particle species can be compared. Figure~\ref{fig:7p7} shows the overlap between the FXT and collider configurations at 7.7 GeV. By considering the difference between the two configurations, it is possible to quantify the systematic error associated with a certain experimental measurement and to extend the measurement to a larger region of phase space.

\section{Conclusion}

The exploration of the phase diagram of strongly interacting matter relies heavily on measurements of heavy-ion collisions at experiments like STAR at RHIC or the future CBM experiment at FAIR.
Based on hints from BES-I that signatures of criticality could be found towards the lower end of RHIC's energy range, a fixed target program building upon several upgrades of the STAR detector was initiated.
eTOF was a key new component. eTOF enlarged the TOF acceptance by 0.7 units in pseudorapidity and provided midrapidity coverage for FXT energies from 4.5 GeV to 7.7 GeV with good charged particle identification. The eTOF system was constructed from prototype detectors and readout components for the future CBM experiment within the FAIR-Phase-0 program and provided a full-scale test stand for these components and their operation. 
108 MRPC counters housed in 36 modules with a total of 6912 readout channels were installed on the east side of the STAR detector and integrated into the STAR trigger, data acquisition, and operation procedures.

In this paper, we have described the details of the system in terms of geometrical layout, acceptance, calibration, hit reconstruction, and particle identification and have demonstrated that the original goals were achieved. A system time resolution of about 70 ps (71.4 ps at 4.5 GeV) and a PID efficiency of about 70\%  was reached with various options for fine tuning efficiency and purity.
This enables eTOF to contribute in an optimized way to analyses with quite different requirements such as the investigation of net-proton number fluctuations, transverse momentum spectra and rapidity density distributions of identified particles, or the search for hyper-nuclei.
A prominent, current investigation being pursued  by the STAR collaboration is the search for a critical point by measuring event-by-event fluctuations of net-baryon or, as a proxy, net-proton numbers. Large event numbers with stable acceptance are required for this search. As explained in this paper, this is a challenge for the free-streaming data acquisition concept employed by CBM that was adapted to triggered data acquisition in eTOF at STAR. A solution has been provided for the fluctuation analysis.  We anticipate  that results on the net-proton number fluctuations, including eTOF data, will be published in the not-too-distant future.

\paragraph{Acknowledgments}
\label{Acknowledgements}
This work was supported in part
by the Office of Nuclear Physics within the U.S. DOE Office of Science, the U.S. National Science Foundation, the Ministry of Science and Technology of China and the Chinese Ministry of Education, National Natural Science Foundation of China, German Bundesministerium f"ur Bildung, Wissenschaft, Forschung and Technologie (BMBF), Helmholtz Association. We gratefully acknowledge the continuous support by the management of STAR, CBM, Brookhaven National Laboratory and Helmholtzzentrum Gesellschaft f\"ur Schwerionenforschung (GSI) within the FAIR Phase-0 program.  

\bibliographystyle{elsarticle-num}
\bibliography{bibliography}

\end{document}